\documentclass[usenatbib]{mn2e}
\newcommand{\bib}{\bibitem[\protect\citeauthoryear}

\usepackage{graphicx}
\usepackage{epsf}
\usepackage{color}
\usepackage{times}

\def\degrees{$^{\circ}$\ }
\def\srm{$\sigma_{\rm RM}~$} 
\def\rmm{$\langle{\rm RM}\rangle~$}

\title[The magnetized medium around the radio galaxy B2\,0755+37]
{
The magnetized medium around the radio galaxy B2\,0755+37:
 an interaction with the intra-group gas
}
\author[D. Guidetti et al.]
   {D. Guidetti\thanks{E-mail: d.guidetti@ira.inaf.it}$^{1,}$$^{2}$,
     R.A. Laing$^2$, J.H. Croston$^3$, A.H. Bridle$^4$,
    P. Parma$^1$\\
    $^1$ INAF -- Istituto di Radioastronomia, via Gobetti 101, I-40129 Bologna,
    Italy \\
    $^2$ European Southern Observatory, Karl-Schwarzschild-Stra\ss e 2, D-85748 
    Garching-bei-M\"unchen, Germany \\ 
    $^3$ School of Physics and Astronomy, University of Southampton,
    Southampton, UK \\
    $^4$ National Radio Astronomy Observatory, Edgemont Road, Charlottesville,
    VA 22903-2475, U.S.A.}

\begin{document}
\date{Accepted. Received; in original form }

\pagerange{\pageref{firstpage}--\pageref{lastpage}} \pubyear{2012}

\maketitle                         

\label{firstpage}
\begin{abstract}

We explore the magneto-ionic environment of the isolated radio galaxy B\,2
0755+37 using detailed imaging of the distributions of Faraday rotation and
depolarization over the radio source from Very Large Array observations at 1385,
1465 and 4860\,MHz and new X-ray data from {\sl XMM-Newton}.  The Rotation
Measure (RM) distribution is complex, with evidence for anisotropic fluctuations
in two regions.  The approaching lobe shows low and uniform RM in an unusual
`stripe' along an extension of the jet axis and a linear gradient transverse
to this axis over its Northern half. The leading edge of the receding lobe shows
arc-like RM structures with sign reversals.  Elsewhere, the RM structures are
reasonably isotropic. The RM power spectra are well described by cut-off power
laws with slopes ranging from 2.1 to 3.2 in different sub-regions. The
corresponding magnetic-field autocorrelation lengths, where well-determined,
range from 0.25 to 1.4\,kpc.  There is little large-scale RM structure in the
receding lobe, where the outer scale of the fluctuations is between 16 and
30\,kpc.  The variation of RM fluctuation amplitude across the source is
inconsistent with that expected from a beta-model gas distribution fitted to our
X-ray data, even
allowing for cavities in the thermal gas associated with the radio lobes.  It is
likely that the fluctuations are mostly produced by compressed gas and field
around the leading edges of the lobes.  The X-ray count rate is too low to
determine the density distribution around 0755+37
in sufficient detail to estimate the field strength accurately, but
values of a few $\mu$G are plausible.  We identify areas of high depolarization
around the jets and inner lobes. These could be produced by dense gas
immediately surrounding the radio emission containing a magnetic field which is
tangled on small scales.  We also identify four ways in which the well known
depolarization (Faraday depth) asymmetry between jetted and counter-jetted lobes
of extended radio sources can be modified by interactions with the surrounding
medium.

\end{abstract}


\section{Introduction}
\label{sec:intro}
Radio continuum observations have detected magnetic fields at microGauss levels
in galaxy disks and halos and in the intergalactic media of groups and clusters
of galaxies.  A detailed knowledge of the strength and structure of these fields
is important to a better understanding of the physical processes in the gaseous
environment of galaxies.  The most direct proof of the presence of magnetic
fields mixed with the intergalactic or intracluster medium (IGM/ICM) is provided
by the detection of diffuse radio synchrotron emission on scales up to Mpc in
galaxy clusters (see e.g. \citealt{Ferrari08} for a review).  
However, the Faraday effect \citep{Faraday} across radio galaxies provides a more
detailed picture of these magnetic fields.

This effect  is the rotation of the plane of polarization of
linearly polarized radiation traveling through a magnetized thermal plasma.
Such a situation occurs for example when
polarized radio galaxies are located behind or embedded in
magnetized intergalactic media.
The rotation $\Delta\Psi$ of the {\bf E}-vector position angle of linearly polarized
radiation by a magnetized thermal plasma is given by:
\begin{equation}
\label{pang}
\Delta\Psi_{[{\rm rad}]}= \Psi(\lambda)_{[{\rm rad}]}~-~\Psi_{0~[{\rm
      rad}]}=\lambda^2_ {[{\rm m}^2]}~{\rm RM}_{[{\rm rad\,m}^{-2}]},
\end{equation}
where $\Psi(\lambda)$ and $\Psi_0$ are the ${\bf E}$-vector position angle of
linearly polarized radiation observed at wavelength $\lambda$ and the intrinsic
angle, respectively.  RM is the {\em rotation measure}.  For a fully resolved
foreground Faraday screen, the $\lambda^2$ relation of Eq.\,\ref{pang} holds
exactly at any observing wavelength.  The RM can be expressed as:
\begin{equation}
{\rm RM}_{[{\rm rad\,m}^{-2}]}=K\int_{0}^{L_{[{\rm kpc}]}}n_{\rm{e}~[{\rm
      cm}^{-3}]}B_{z~[\mu{\rm G}]}dz_{[{\rm kpc}]}\,,
\label{equarm}
\end{equation}
where $n_e$ is the electron number density in the thermal plasma, $B_{z}$ is the magnetic
field along the line-of-sight and $L$ is the integration path. $K = 811.9$ for
the units in equation~\ref{equarm}.

Observations of Faraday rotation variations across extended radio galaxies, once
corrected for the contribution from our Galaxy, allow us to derive information
about the integral of the density-weighted line-of-sight field component. In
several cases, the linearity of the relation between $\Delta\Psi$ and
$\lambda^2$ over a range $|\Delta\Psi| \ga 45^\circ$ and the absence of
depolarization at high angular resolution require that the majority of the
material responsible for the variations must be in front of the synchrotron
emission (e.g.\ \citealt{cyga,laing08,Guidetti10}); most observations of other sources are
also consistent with pure foreground rotation. If the Faraday rotation is produced by
hot thermal plasma, the density distribution can be derived from X-ray
observations and the magnetic-field strength can be estimated. The main
remaining uncertainties are the field topology and the geometry of the source
and plasma.


There has been some controversy about the relative importance of Faraday
rotation from material which has been affected by the radio galaxy (e.g.\
compressed by a bow-shock or entrained in a turbulent mixing layer) and the
undisturbed intra-group or intra-cluster medium (e.g.\ \citealt{RB03,Ens03}).
Many of the published RM images of radio galaxies show patchy patterns without
any obvious preferred direction in their iso-RM contours which might indicate a
correlation with the source structure. In these cases, the hypothesis that the
RM is dominated by the undisturbed hot component is self-consistent.  Models in
which the magnetic field is an isotropic, random variable with structure on a
wide range of scales and with a strength that scales with density can fit the
observations for reasonable source geometries
(e.g. \citealt{Murgia,Guidetti08,laing08,Kuchar11,Guidetti10}).  Three of the
examples discussed in these references, Hydra\,A, 3C\,31 and 3C\,449, are
central radio galaxies in groups or clusters with {\em tailed} radio
morphologies.

In contrast, we have demonstrated unequivocally that the interaction between
radio galaxies and their immediate environments can affect the magnetization of
the plasma surrounding some {\em lobed} radio galaxies \citep{Guidetti11}.
These sources show areas of ordered rotation (`RM bands') with iso-RM contours
orthogonal to the axes of the radio lobes.  They are embedded in different
environments, but all show evidence of strong interaction with the surrounding
medium, either from brightness gradients at their leading edges or from
cavities and shells of swept-up and compressed material observed in
X-rays. Enhanced Faraday rotation should occur in the shocked plasma surrounding
a radio lobe which is expanding supersonically and is indeed inferred to be
present in Cygnus\,A \citep*{bowshock}. If the field in the undisturbed ambient
plasma is disordered, then the Faraday rotation produced by the post-shock
plasma will still appear patchy, as it does in the simulations by
\citet*{huarte}.  The anisotropy of the RM bands was therefore unexpected: a
two-dimensional magnetic field draped around the front end of the radio lobes is
required to produce the observed structures \citep{Guidetti11}.


\citet{Guidetti11} also presented evidence for significant {\em depolarization}
along the edges of the inner lobes in two sources, M\,84 and 3C\,270, coincident
with shells of enhanced X-ray emission. There is little sign of associated
large-scale variations of RM, and the most likely explanation is that the field
is tangled on small scales.

The picture that emerges is that the magneto-ionic environments of radio
galaxies are complex,  potentially having at least three components: 
the undisturbed IGM/ICM; a region of enhanced density and (sometimes highly anisotropic) field
around the leading edges of active lobes, and smaller-scale field fluctuations
in denser thermal gas close to the nucleus.

In order to distinguish between these (and other) components, we need to image the
Faraday rotation and depolarization at a large number of independent points with
high signal-to-noise ratio, allowing us to estimate spatial statistics
accurately, as well as to determine large-scale variations.  Highly-polarized
radio galaxies with large lobes are suitable targets, and in this paper we
present observations of one of these: B2\,0755+37\footnote{From now on we drop
the B2.}, located at the centre of a
nearby and very sparse group of galaxies.  We have imaged the RM and
depolarization over the radio source
at two angular resolutions using high-quality VLA data in the 20
and 6\,cm bands, revealing another good (but different) example
of anisotropic RM structure. In this paper we give a detailed analysis of the RM
fluctuations across the source. We also present new {\sl XMM-Newton} X-ray 
observations and use them to characterize the hot gas associated with the galaxy
group.

The paper is organized as follows.  Section~\ref{radio_opt} summarizes the
properties of the host galaxy and group and the main features of the
polarized radio emission of 0755+37. In Section~\ref{07x}, we describe our new {\sl XMM-Newton}
observations and use them to derive the density profile and temperature of the
group gas.  Section~\ref{2d} presents
high-quality RM and depolarization images.  The spatial statistics of RM are
analysed using structure-function techniques and depolarization simulations in
Section~\ref{presfunc}.  Section~\ref{discuss} discusses the implications of
our results for the interaction of the source with the surrounding magnetized
medium.  Finally, Section~\ref{concl} summarizes our results.

\begin{figure*}
\centering
\includegraphics[width=8.5cm]{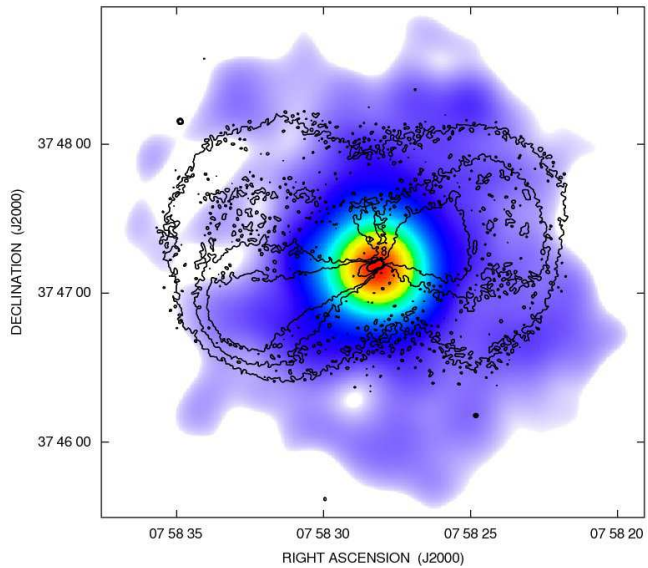}
\caption[]{Overlay of the X-ray emission detected by {\sl XMM-Newton} around
  0755+37 (colour) on contours of 1385-MHz emission at a resolution of
  1.3\,arcsec. The X-ray image has been smoothed with a Gaussian function of
  FWHM 14.1\,arcsec to give a resolution of $\approx$15.3\,arcsec FWHM.  A point source at
  RA 07$^{\rm h}$ 58$^{\rm m}$ 26$^{\rm s}$, Dec +37\degr 46\arcmin 26\arcsec
  (coincident with a known quasar) has been subtracted from the X-ray image.
\label{0755X}}
\end{figure*}

\section{0755+37 and its gaseous environment}
\label{07intro}

\subsection{Radio and optical properties}
\label{radio_opt}

0755+37 is an extended FR\,I radio source \citep{Fana74} whose optical
counterpart, NGC\,2484, is a D-galaxy with a redshift of 0.04284 \citep{Falco99}.
We assume a cosmology with $H_0$ = 71 km s$^{-1}$Mpc$^{-1}$, $\Omega_M$ = 0.3,
and $\Omega_{\Lambda}$ = 0.7, implying that 1\,arcsec corresponds to 0.833\,kpc.
The source is one of the most isolated radio galaxies in the B2 sample and is
classified as the central member of a very poor group \citep{Mul03}.

The VLA observations in the 20 and 6\,cm bands used in this paper were presented
by \citet{laing11}. At a resolution of 1.3\,arcsec the source shows a very
bright core and two-sided jets embedded in diffuse lobes of lower surface
brightness.  The jet East\footnote{We refer to the E (East) and W (West) radio
lobes and jets, following \citet{laing11}.} of the nucleus is the brighter one and the
inclination of the jets to the line of sight is estimated to be about
35\degrees with the E side approaching (\citealt{Bon00}; Laing \& Bridle, in
preparation).  Although the lobes resemble those in powerful FR\,II sources in
morphology, spectral-index distribution and magnetic-field structure, they do
not show evidence for hot-spots or shocks at the ends of the jets
\citep{laing11}.  The radio emission is highly polarized with the apparent
magnetic field aligned perpendicular to the highest brightness gradients over
both lobes.

\begin{figure}
\centering
\includegraphics[width=6.5cm]{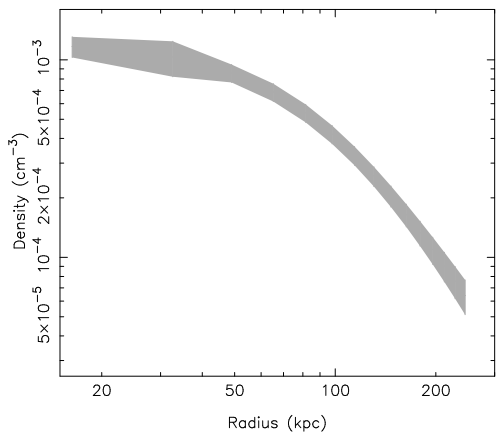}
\caption{Radial profile of proton density for the group-scale hot gas component
  derived by fitting a beta model to the {\sl XMM-Newton} data as described in
  the text. The grey-shaded area indicates the $1\sigma$ uncertainty, which
  takes into account uncertainties in all the model parameters and in the
  temperature measurement.
\label{xmmdensprof}}
\end{figure}

\subsection{{\sl XMM-Newton} observations and X-ray properties}
\label{07x}

0755+37 was previously observed with the PSPC and HRI instruments on {\sl ROSAT}
  \citep{Worr94, Worr00} and with {\sl Chandra} \citep{Worr01}.
The X-ray emission consists of thermal and non-thermal components, respectively
associated with the hot medium on group and galactic scales, and with the core
and brighter jet \citep{Worr00,Worr01,Wu07}. The X-ray surface brightness
profile for the thermal emission is spatially resolved and can be modelled as
the superposition of two components: a dense, kpc-scale corona seen by {\sl Chandra}
\citep{Worr01} and a much larger and more tenuous group atmosphere detected by
the {\sl ROSAT} PSPC \citep{Worr00}.

\begin{figure*}
\centering
\hspace{10pt}
\includegraphics[width=17cm]{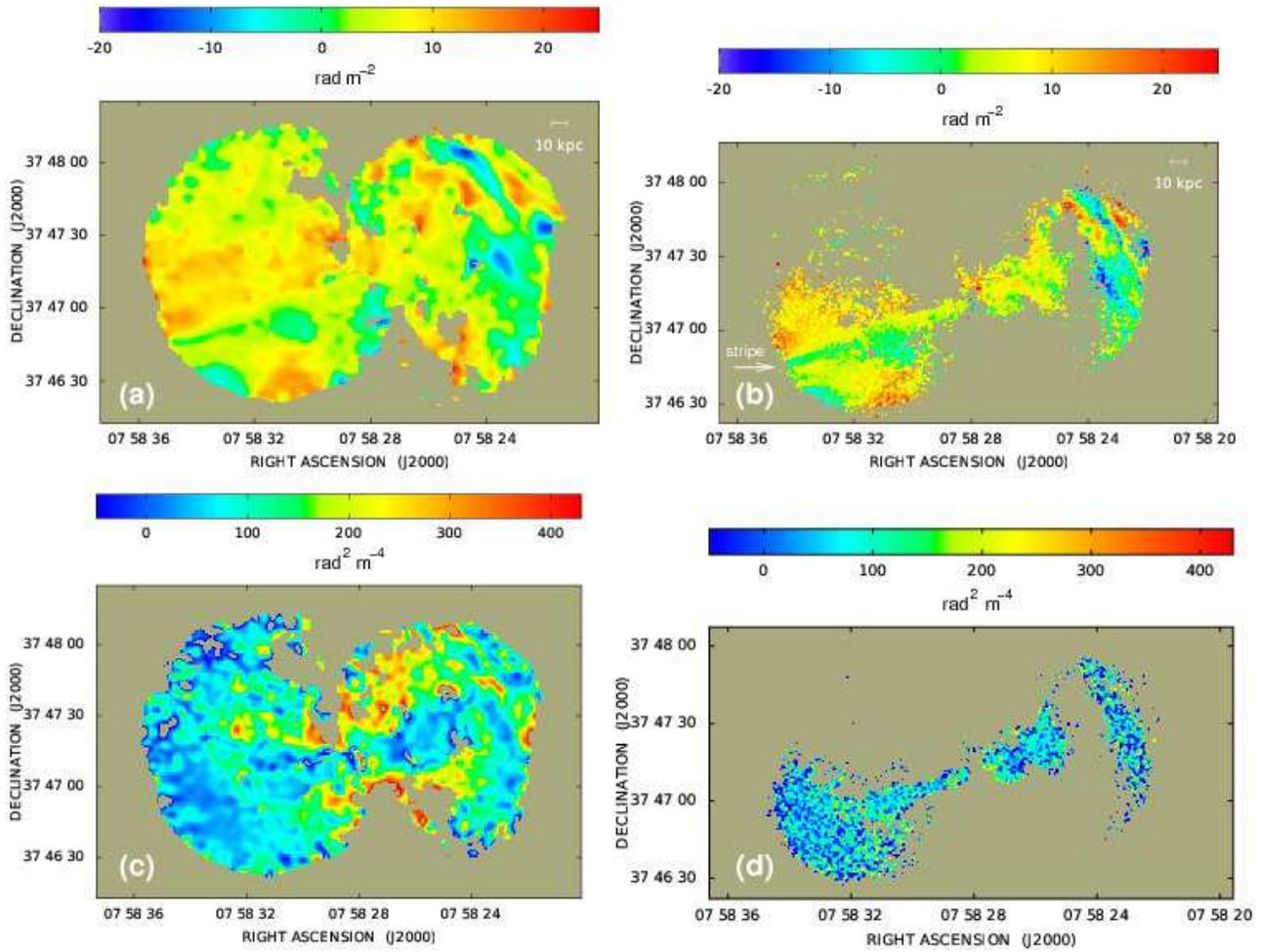}
\caption[]{(a) and (b): images of rotation measure, computed from
weighted least-squares fits to the position angle -- $\lambda^2$ relation at 
frequencies of 1385.1, 1469.1 and 4860.1\,MHz. (c) and (d): images of Burn law $k$ computed
from weighted least-squares fits to the relation 
$\ln p(\lambda)=\ln p(0)-k\lambda^4$ at the same  
frequencies. The resolutions are: (a) and (c) 4.0\,arcsec
FWHM; (b) and (d) 1.3\,arcsec FWHM.
\label{0755RMkfig}
}
\end{figure*}

\begin{table}
\small\addtolength{\tabcolsep}{-4pt}
\caption{Parameters of the hot gas distribution as deduced from {\sl Chandra}
  and {\sl XMM-Newton} observations \citep[this paper, Section~\ref{07x}]{Worr01}. Col. 1: component;
Col. 2: temperature (keV); 
Cols. 3, 4, 5: best-fitting core radius (kpc), central proton density and $\beta$
parameter, respectively. 
Col. 6: reference.
}
\vspace{3pt}
\centering
\begin{tabular} {@{}c c c c c c}  
\hline
Component & $kT$ &  $r_{c}$ & $n_{0}$    & $\beta$ & ref. \\
          & keV &  kpc       &  cm$^{-3}$ &         &      \\
\hline
&&&&&\\  
Corona & 0.84$^{+0.07}_{-0.16}$  & 1.7 &  9.9~$\times$~10$^{-2}$  & 0.8 & 1 \\
Group  & 1.3 $^{+0.2 }_{-0.1 }$  &  66 &  1.2~$\times$~10$^{-3}$  & 0.7 & 2  \\      
&&&&&\\  
\hline
\multicolumn{1}{l}{\scriptsize References:}&\multicolumn{5}{l}{\scriptsize (1) \citet{Worr01}} \\
                                           &\multicolumn{5}{l}{\scriptsize (2) Section~\ref{07x}}\\
\label{ongoingX}
\end{tabular}
\end{table}

0755+37 was observed with {\sl XMM-Newton} on 2002 March 27 for $\sim 65$
ks. Unfortunately the observation was affected by high background levels, which
meant that the EPIC pn data were unusable. The EPIC MOS data were processed in
the standard manner using {\sl XMM-Newton} SAS version 11.0, and the latest
calibration files from the {\sl XMM-Newton} website. The data were filtered
according to the standard flag and pattern masks (PATTERN $\leq 12$ and
\#XMMEA\_EM, excluding bad columns and rows). Good-time-interval (GTI) filtering
was applied to remove periods of high background, which led to filtered exposure
times of 51.3 and 50.0 ks for the MOS1 and MOS2 cameras, respectively. Images
were extracted in the 0.5 -- 5 keV energy range, and combined. They
were not vignetting-corrected to minimize artefacts associated with the particle
background.  A confusing point source at RA 07$^{\rm h}$ 58$^{\rm m}$ 26$^{\rm
s}$, Dec +37\degr 46\arcmin 26\arcsec (J2000), coincident to within {\sl
XMM-Newton} positional errors with a known quasar, was subtracted from the X-ray
image.

Fig.~\ref{0755X} shows an overlay of the {\sl XMM-Newton} image smoothed with a
Gaussian of 14.1\,arcsec FWHM on contours of 1385-MHz emission at a resolution of
1.3\,arcsec FWHM. The {\sl XMM-Newton} image shows an apparent deficit in X-ray
surface brightness associated with the Northern part of the E radio lobe. We
investigated the significance of this feature by dividing the radio source
into four quadrants (using the jet to define an axis) and found that the deficit
is significant only at the 2$\sigma$ level. Detection of any cavities associated
with the radio lobes would require significantly deeper X-ray imaging.

Spectral and spatial analysis were carried out on the filtered X-ray events
files using the methods of \citet{Cro08}. These included double background
subtraction using filter-wheel closed files to account for the particle
contribution and local reference regions to model the Galactic and cosmic X-ray
emission. Results were compared with those using a local background only in a
restricted energy band (0.3 -- 2.0 keV) to ensure that background contamination
was not affecting our measurements.

A global spectrum was obtained for an annular region between radii of 60 and 300
arcsec (excluding the central AGN). This was fit with an {\sc apec} model. We
found a best-fitting temperature of $kT = 1.3^{+0.2}_{-0.1}$\,keV for a fixed
abundance of 0.3 times solar. A surface brightness profile was extracted from
the MOS1 and MOS2 datasets, and the combined profile was fit with a beta model
convolved with the {\sl XMM-Newton} point-source function using the Markov-Chain
Monte Carlo fitting methods described in \citet[an updated version of the method
described in \citealt{Cro08}]{Goodger12}. The emission from the nucleus, inner
jet and corona all lie within the excluded inner circular region.  The beta model
parameters are individually poorly determined, with Bayesian estimates of $\beta
= 0.7$ and $r_{c} = 79$\,arcsec (66\,kpc). Nevertheless, the density profile
(shown in Fig.~\ref{xmmdensprof}) is well constrained. We estimated the total
(unabsorbed) bolometric luminosity to be $1.5^{+0.1}_{-0.2} \times
10^{42}$\,erg\,s$^{-1}$ by integration of the surface brightness profile. We
cannot entirely rule out the possibility that the high particle background in
this dataset has affected our temperature estimate; however, we believe that the
group temperature is indeed higher than \citet{Worr00}'s estimate of
0.73\,keV. Our measured temperature is high for the group's luminosity (e.g.\
based on the X-ray luminosity--temperature relations discussed in
\citealt{Cro08}), but within the scatter for other radio-loud groups. The value
measured by \citet{Worr00} is more consistent with the luminosity-temperature
relationship for radio-quiet groups, which makes it slightly low compared to
other radio-galaxy environments of similar luminosity.  The density profile,
which is of most importance for comparison with the rotation measure results, is
not significantly affected by the uncertainty in temperature and our luminosity
estimate is consistent with that of \citet{Worr00}.

\section{Rotation measure and depolarization distributions} 
\label{2d}

\subsection{Rotation measure images}
\label{ongoingRM}

We produced images of RM and its rms error by weighted least-squares fitting to
the polarization angle maps $\Psi(\lambda)$ as a function of $\lambda^2$
(Eq.\,\ref{pang}) at frequencies of 1385.1, 1464.9 and 4860.1\,MHz.  The RM
images were made at resolutions of 1.3 and 4.0\,arcsec FWHM and calculated only
at pixels with polarization angle uncertainties $\leq$10\,\degrees at all
frequencies. They are shown in Figs~\ref{0755RMkfig}(a) and (b).

The fits are very good at both resolutions, showing no evidence for  deviations from
a $\lambda^2$ law, although the total range of rotation is quite small. In what
follows, we implicitly assume that there is no small-scale mixing of thermal and
relativistic plasma, so that all of the observed Faraday rotation is due to
foreground thermal material. This is consistent with our observations of
0755+37, but is not explicitly required by them. We argue by analogy with
FR\,I sources in denser environments  (for which foreground rotation is definitely
required; e.g.\ \citealt{laing08,Guidetti10}) that this is also the case for 0755+37.

At 4.0\,arcsec resolution, the RM distribution has \rmm $= +5.2$\,rad\,m$^{-2}$
with \srm = 4.9\,rad\,m$^{-2}$ and an average fitting error $\sigma_{\rm
  RM_{fit}} = 0.8$\,rad\,m$^{-2}$.  At 1.3\,arcsec the mean RM is
$+5.0$\,rad\,m$^{-2}$ with \srm=5.9\,rad\,m$^{-2}$ and $\sigma_{\rm RM_{fit}} =
2.0$\,rad\,m$^{-2}$. The range in RM at the higher resolution is from $-20$ to
$+25$\,rad\,m$^{-2}$.

\begin{figure}
\includegraphics[width=5cm]{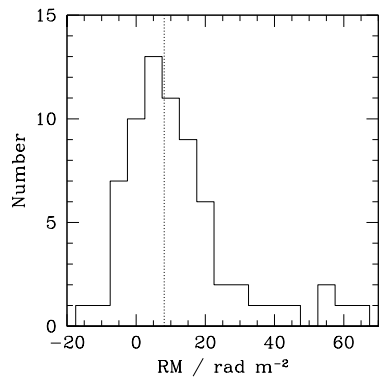}
\caption[]{Histogram of RM's for the 69 sources within $4^\circ$ of 0755+37 with 
  measurements tabulated by \citet{NVSSRM}. The median value,
  $+8.1$\,rad\,m$^{-2}$, is indicated by the vertical dotted line.\label{NVSSfig}
}
\end{figure}

In order to estimate the contribution from Galactic Faraday rotation, we used
measurements of RM for compact sources selected from the NVSS Catalogue
\citep*{NVSSRM}. The RM distribution for the 69 sources within a radius of
4$^\circ$ of 0755+37 is shown in Fig.~\ref{NVSSfig}. This has a well-defined
peak centred on a small positive value. The median of the distribution is
$+8.1$\,rad\,m$^{-2}$ and the error-weighted mean is $+7.8 \pm 0.6$\,rad\,m$^{-2}$. We
adopt the former as the best available estimate of the Galactic RM.

\begin{figure}
\includegraphics[width=8cm]{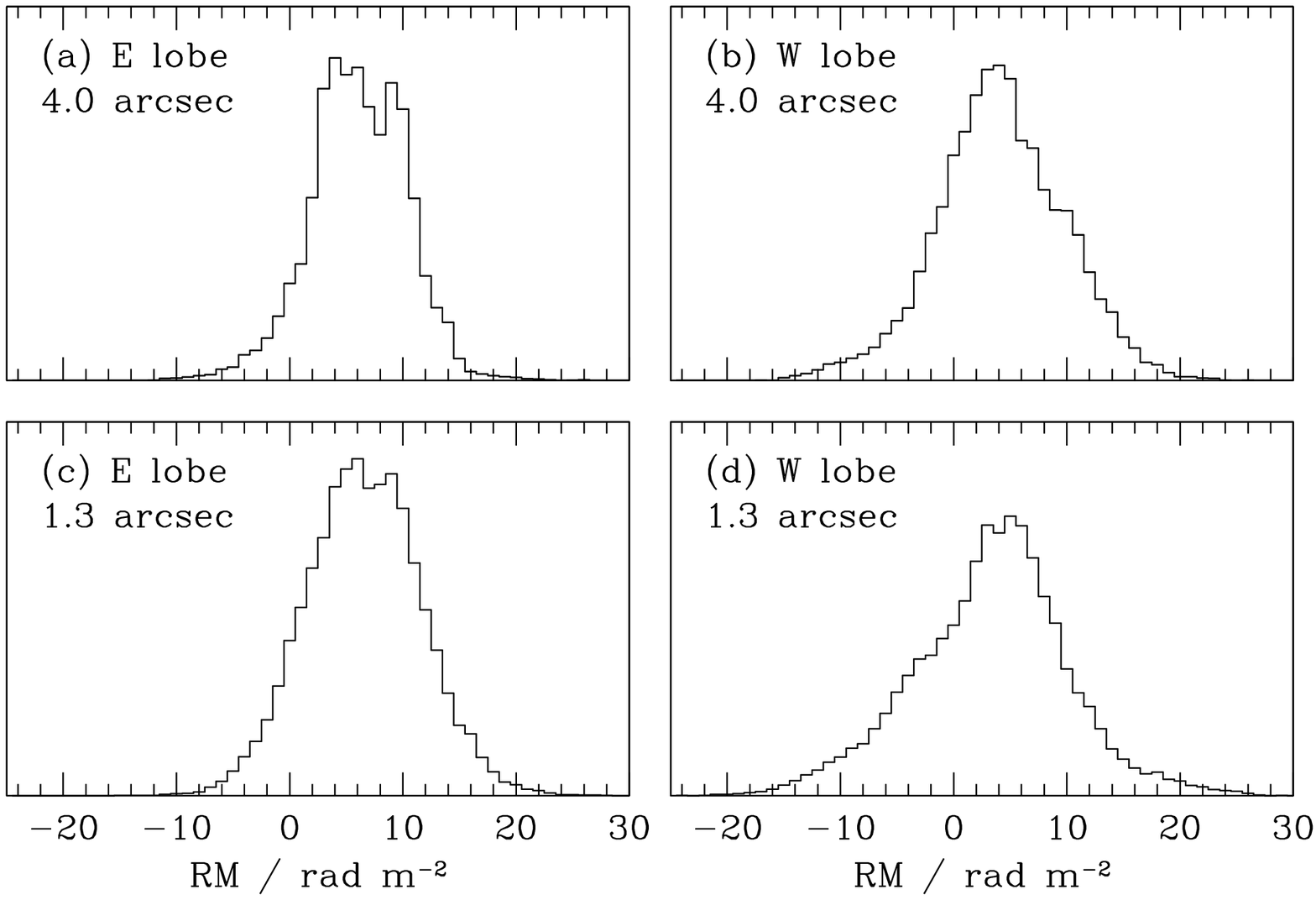}
\caption[]{Histograms of RM over the E and W lobes of 0755+37. (a) E lobe,
  4.0\,arcsec FWHM. (b) W lobe 4.0\,arcsec FWHM. (c) E lobe, 1.3\,arcsec
  FWHM. (d) W lobe, 1.3\,arcsec FWHM.\label{fig:rmhist}
}
\end{figure}

The W lobe shows a higher average
fluctuation level, with \srm = 5.6\,rad\,m$^{-2}$ and \srm = 6.6\,rad\,m$^{-2}$ at
4.0 and 1.3-arcsec resolutions, respectively. The equivalent numbers for the E
lobe are \srm = 4.1\,rad\,m$^{-2}$ and 4.9\,rad\,m$^{-2}$. Fig.~\ref{fig:rmhist}
shows histograms of the RM distributions over the E and W lobes separately at
the two resolutions. The larger range in the W lobe is evident. 

\begin{figure*}
\centering
\includegraphics[width=13cm]{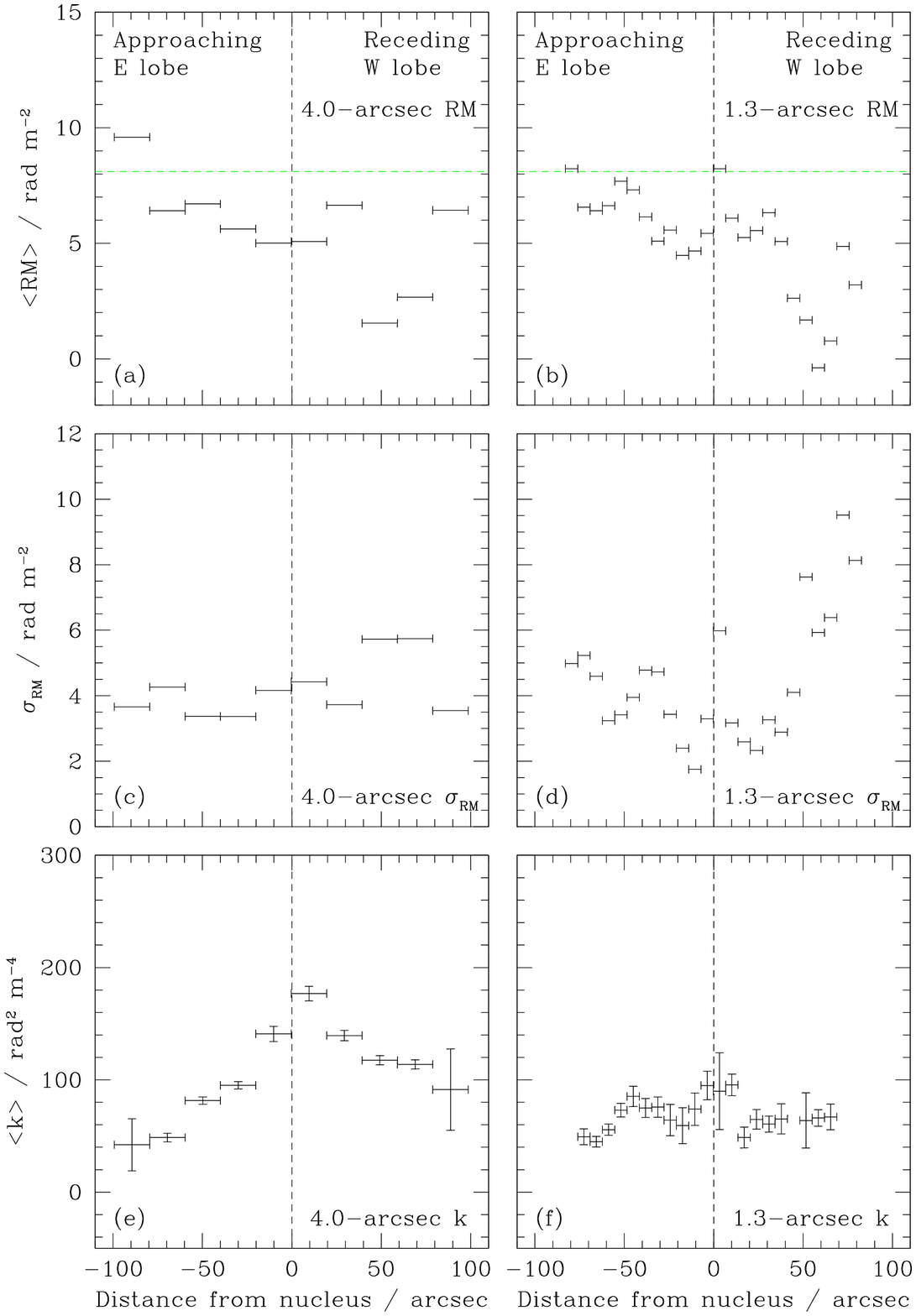}
\caption[]{Average profiles of \rmm, \srm and Burn law $k$ for 0755+37 parallel
  to the major axis of the source (in PA $-70^\circ$) 
  plotted against distance from the nucleus. (a) and (b) \rmm. The dashed green
  horizontal line is our best estimate of the Galactic contribution. (c) and (d)
  \srm, corrected to first order for fitting errors. (e) and (f) $k$. The error
  bars on the $k$ profiles are estimated from the uncertainties on the fits. 
  The angular resolutions are 4.0 (a, c, e) and 1.3\,arcsec FWHM (b, d, f). The
  data were averaged in boxes of length 19.8\,arcsec and 6.9\,arcsec, respectively,
  at the two resolutions, extended in width to cover all unblanked pixels.

\label{0755kprof}
}
\end{figure*}

There is an asymmetry in RM structure between the E and W lobes at both
resolutions.  To a first approximation, the N part of the E (approaching) lobe
is characterized by a smoothly-varying, positive RM whereas the S part of that
lobe and the whole of the W lobe show irregular fluctuations.  The largest RM
fluctuations, of both signs, are at the end of the W lobe.  Profiles of \rmm
averaged in boxes along the source axis (taken to be in position angle
$-70^\circ$) at the two resolutions are shown in Figs~\ref{0755kprof}(a) and
(b). There is a trend to lower values of RM from E to W and the dispersion
between the boxes is larger in the W lobe. The corresponding profiles of \srm,
corrected to first order for fitting error, are shown in Figs
~\ref{0755kprof}(c) and (d). There are three local maxima in the
higher-resolution profile: at the outer edges of both lobes and close to the
nucleus.

\begin{figure*}
\includegraphics[height=7cm]{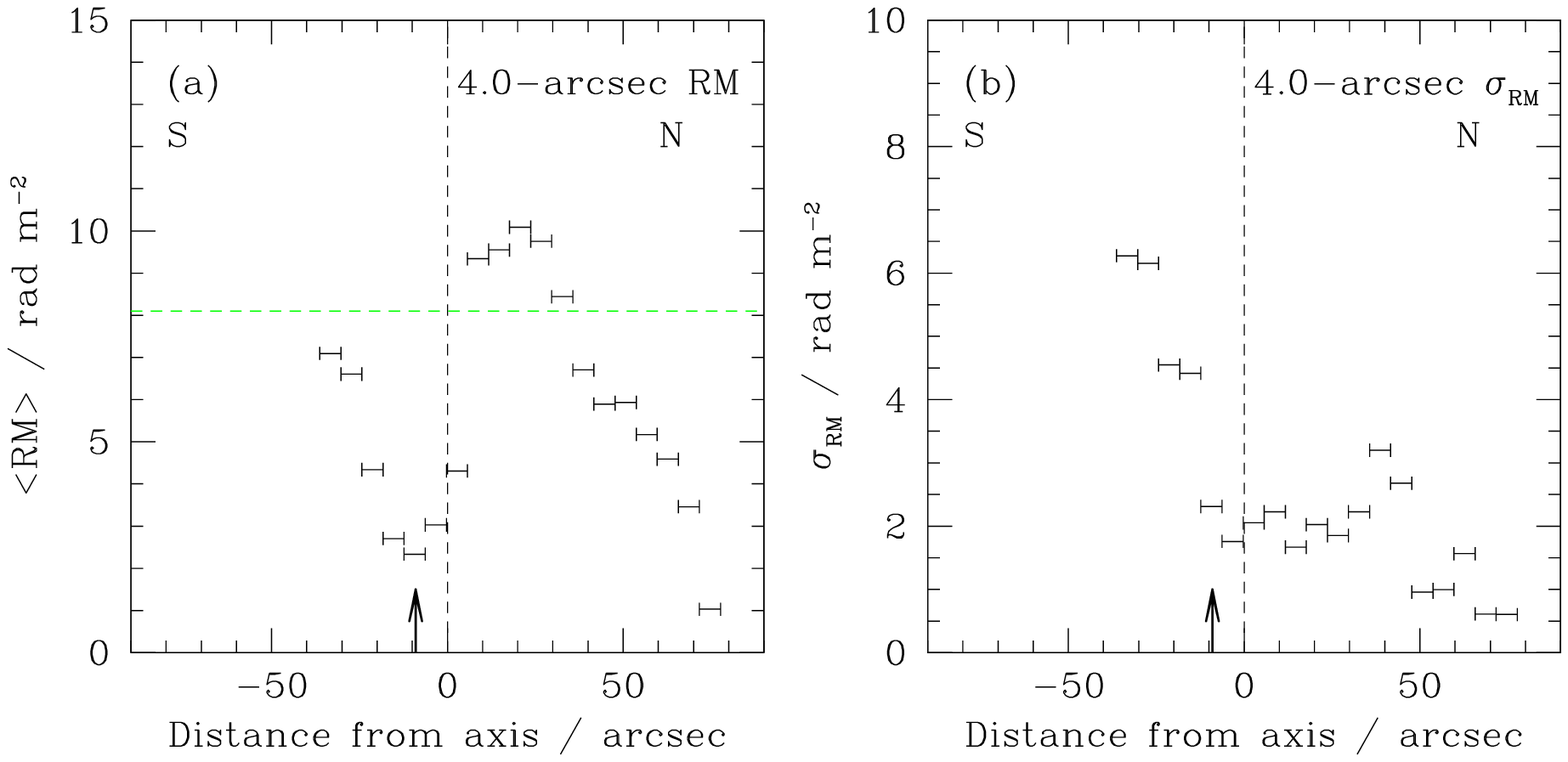}
\caption{Profiles of \rmm and \srm  perpendicular to the source axis in the E
  lobe plotted against distance from the lobe axis (defined as a line in PA
  $-70^\circ$ passing through the core).
The angular resolution is 4.0\,arcsec FWHM. Data were averaged in boxes of
length 6\,arcsec transverse to the source axis (in PA 20$^\circ$), extended to
cover all of the unblanked pixels in the lobe. The vertical arrow shows the
location of the `RM stripe' discussed in the text. (a) \rmm. The estimated Galactic
contribution is indicated by the horizontal green dashed line. (b) \srm. The
values have been corrected to first order for the fitting error.
\label{tran_prof}
}
\end{figure*}

Profiles of \rmm and \srm transverse to the source axis in the E lobe at
4.0-arcsec resolution are plotted in Fig.~\ref{tran_prof}.  The N half of the
lobe is characterized by a linear gradient of \rmm and a low level of random
fluctuation.  \rmm decreases from $+10$\,rad\,m$^{-2}$ on-axis to
$+1$\,rad\,m$^{-2}$ at the edge of the emission. In order to quantify this
gradient, we fit the relevant portion of the RM image with a function of the
form:
\[
{\rm RM}(x,y) = {\rm RM}_0 + ax + by
\]
where $x$ is along the source axis increasing towards the nucleus (PA $-70^\circ$)
and $y$ orthogonal to it, increasing in PA $+20^\circ$.  We found $a =
-0.032$\,rad\,m$^{-2}$\,arcsec$^{-1}$ and $b =
-0.111$\,rad\,m$^{-2}$\,arcsec$^{-1}$, so the direction of steepest gradient is in
PA $+36^\circ$, not quite orthogonal to the source axis. The residual
fluctuation levels before and after subtraction of the best-fitting linear
gradient are 3.5 and 2.6\,rad\,m$^{-2}$ respectively, compared with a mean
fitting error of 1.3\,rad\,m$^{-2}$.

\begin{figure}
\includegraphics[width=6.5cm]{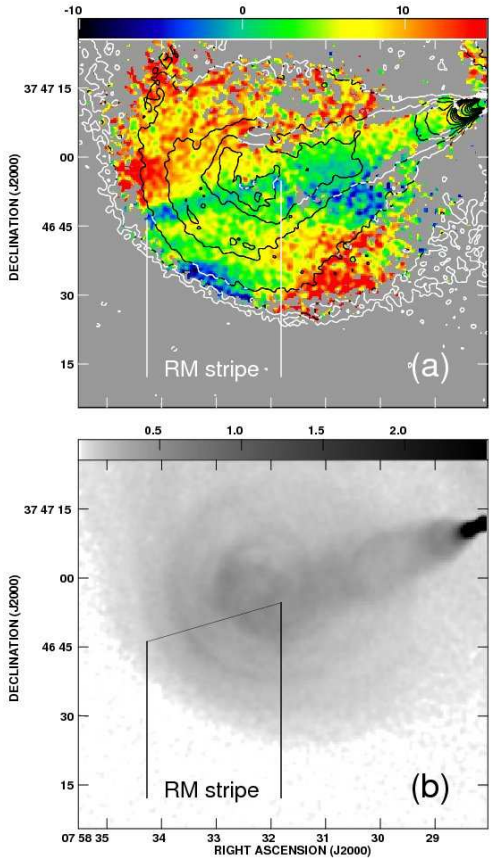}
\caption{(a) The RM of the E lobe of 0755+37 at a resolution of 1.3\,arcsec,
  with a colour range chosen to emphasis the narrow stripe of near-zero RM
  described in the text. (b) 4860-MHz total intensity at the same
  resolution, also showing the location of the RM stripe.
\label{rmstripe}
}
\end{figure}

The area of smoothly-varying, positive RM in the E lobe is bounded by a thin,
straight structure elongated parallel to the source axis with almost zero \rmm.
This `stripe' is  striking at 1.3-arcsec resolution
(Fig.~\ref{rmstripe}a). It is almost but not quite aligned with the nucleus
and lies along a continuation of the main jet axis. It starts just before the
apparent termination of the jet and continues as far as the boundary of the lobe,
in the process crossing the prominent circular ridges of emission
(Fig.~\ref{rmstripe}b).  It does not obviously extend over the
high-brightness emission of the E jet closer to the nucleus, although \rmm
remains small there. Correction for the Galactic contribution implies that the
intrinsic RM of the stripe is negative ($\approx -6$\,rad\,m$^{-2}$).  S of the
stripe, \rmm again becomes positive, and the fluctuation level increases away
from the source axis.

\begin{figure}
\includegraphics[width=7.5cm]{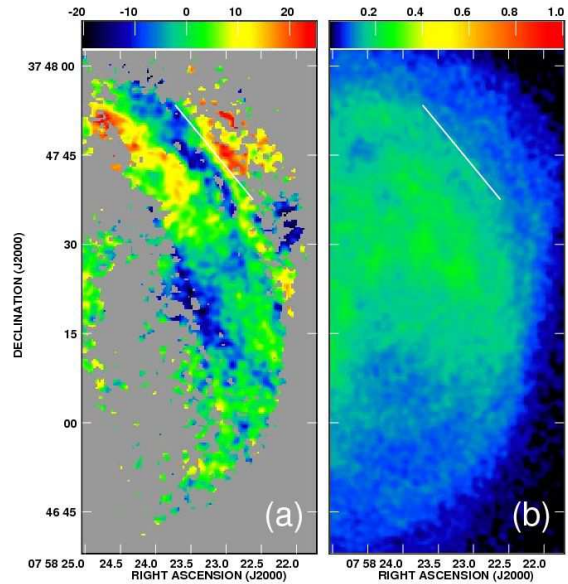}
\caption{(a) The RM of the end of the W lobe of 0755+37 at a resolution of
  1.3\,arcsec. The colour range is $-$20 to +25\,rad\,m$^{-2}$. 
(b) 4860-MHz total intensity at the same
  resolution with a colour range from 0.01 to 1\,mJy\,beam$^{-1}$. The
  brightness step which coincides with a change in sign of RM is marked by a
  white line on both plots.
\label{fig:wlobe}
}
\end{figure}

In contrast, the outer half of the W lobe has the highest level of RM
fluctuations in the source. The iso-RM contours have a preferred direction in
position angle $\approx$50$^\circ$, so the field responsible for the Faraday
rotation cannot be isotropic.  The clearest example of this anisotropy is the
change in sign of RM at the prominent brightness step marked on
Fig.~\ref{fig:wlobe}.  The RM fluctuations in this part of the source have
alternating signs with respect to the Galactic value and must therefore be
associated with field reversals.  They resemble the RM bands discussed by
\citet{Guidetti11}, but are neither as closely orthogonal to the source axis nor as well-defined as the clearest examples of bands.

\subsection{Depolarization}
\label{07dp}

We have chosen to parametrize the dependence of the degree of polarization
$p(\lambda)$ on wavelength by the Burn law \citep{Burn66}:
\begin{equation}
\label{eq:dp}
p(\lambda)=p(0)\exp(-k\lambda^4),
\end{equation}
where $p(0)$ is the intrinsic value of the degree of polarization.  This
expression is appropriate for well (but not fully) resolved foreground rotation:
if the spatial variations of RM can be approximated as linear across the
(Gaussian) beam, then the wavelength dependence of the depolarization is
expected to follow the relation with $k$=2$\arrowvert\nabla{RM}\arrowvert^2
\sigma^2$ and $\sigma = (8\ln 2)^{-1/2} {\rm FWHM}$ \citep{laing08}. The
position angle continues to obey the $\lambda^2$ relation of Equation~\ref{pang}
in this approximation. Simulations of foreground rotation by magnetic fields
with more complex power spectra \citep{laing08} 
show that Equation~\ref{eq:dp} is a good approximation to the mean value 
of $p(\lambda)$ in the short-wavelength
limit ($p(\lambda)/p(0) \la 0.5$).

Equation~\ref{eq:dp} is consistent with our observations of 0755+37 and fitting
to it allows us to include all three frequencies in the analysis.  We therefore
produced Burn law $k$ images at 4.0 and 1.3-arcsec resolution, just as for the
RM, by fitting to the relation $\ln p(\lambda) = \ln
p(0) - k\lambda^4$ at all pixels with $p>4\sigma_p$ at all three
frequencies.
The Burn law $k$ maps are shown in Figs\,\ref{0755RMkfig}(c) and (d).  At 4.0
and 1.3-arcsec resolutions, the mean values of $k$ are 108\,rad$^2$m$^{-4}$ and
63\,rad$^2$m$^{-4}$ respectively, corresponding to depolarizations
$p(\lambda)/p(0) = 0.79$ and 0.87 at our lowest frequency.  One obvious 
reason for this difference is that RM fluctuations on intermediate scales cause
depolarization at the lower resolution: this is the primary effect in the
brighter parts of the source. In addition, some of the low-surface-brightness
emission, which can be detected in polarization only at 4.0-arcsec resolution,
shows more rapid depolarization.   

\begin{figure}
\centering
\includegraphics[width=8.5cm]{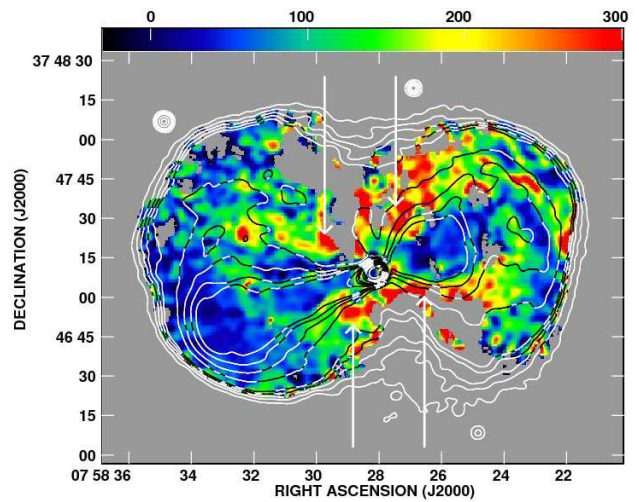}
\caption{Superposition of a false-colour image of Burn law $k$ on contours of
  total intensity at 4860\,MHz. The resolution is 4.0\,arcsec FWHM. The V-shaped
  regions of high depolarization described in the text are marked by arrows.
\label{kI}}
\end{figure}

The most depolarized parts of the source are in the W lobe, in a V-shaped region
with $\langle k \rangle \approx 240$\,rad$^2$m$^{-4}$ on either side of the
counter-jet. This is most clearly seen on Fig.~\ref{kI}, where the 4.0-arcsec
$k$ image is superposed on contours of total intensity at the same
resolution. There are also traces of a similar but less prominent structure in
the W lobe, around the main jet. The cap of emission at the end of the W lobe is
also more depolarized ($\langle k \rangle = 109$\,rad$^2$m$^{-4}$) than its
eastern counterpart ($\langle k \rangle = 45$\,rad$^2$m$^{-4}$).  The
combination of these features leads to a marked asymmetry, illustrated by a
longitudinal profile of $\langle k \rangle$ averaged transverse to the source
axis (Fig.~\ref{0755kprof}e). $\langle k \rangle = 87$\,rad$^2$m$^{-4}$ and
137\,rad$^2$m$^{-4}$ for the whole of the E and W lobes, respectively.

We note that much of the depolarized emission is well inside the edges of the
source and is resolved: i.e.\ the depolarization is not observed just at
individual pixels at the boundaries of the emission where the signal to noise is
low.  In addition, simulations of our fitting procedure \citep{laing08} confirm
that our blanking criterion gives secure detections of depolarization with
negligible bias, so we believe that even the high depolarizations seen at the
edges of the source are real. There are many independent points with significant
detections of high depolarization: for example, the depolarized region to the N
of the counter-jet has $\sim$2000 pixels (40 independent beam areas). Over this
region, the detection of depolarization is highly significant: $\langle k
\rangle = 220$\,rad$^2$\,m$^{-4}$ compared with a mean error derived from the
fits of 40\,rad$^2$\,m$^{-4}$. Finally, high depolarization is not observed
close to the boundary of the E lobe, where it might be expected if it is an edge
artefact. We are therefore confident in the reality of the highly depolarized
regions.

At 1.3-arcsec resolution (Fig.~\ref{0755RMkfig}d), only the jets and the caps
of the two lobes have enough polarized emission to be imaged above the noise level at any of
the three frequencies and we therefore cannot determine $k$ for the V-shaped
regions around the jets. The cap of the W lobe has $\langle k \rangle =
63$\,rad$^2$m$^{-4}$ compared with 109\,rad$^2$m$^{-4}$ at 4.0-arcsec
resolution, indicating that the depolarization at the lower resolution is partly
due to fluctuations in RM on scales between 1 and 4\,arcsec. There is much less
difference between the values of $\langle k \rangle$ for the cap of the E lobe:
these are 45 and 42\,rad$^2$m$^{-4}$, respectively at 4.0 and 1.3-arcsec
resolution. The combination of these effects causes the longitudinal profile of
$k$ at 1.3-arcsec resolution (Fig.~\ref{0755kprof}f) to be much flatter and
more symmetrical than the equivalent at lower resolution.

\subsection{Structure functions}
\label{presfunc}

To quantify the two-dimensional fluctuations of Faraday rotation measure
on scales larger than the observing beam, we used the RM structure function
(\citealt*{SCS}; \citealt{MS96}) defined by:
\begin{equation}
S(r_\perp)=\rm{<[RM({\bf r}_\perp + {\bf r}_\perp^\prime)-RM({\bf r}_\perp^\prime)]^2>}  
\label{sfunction}
\end{equation}
where ${\bf r_{\perp}}$ and ${\bf r'_{\perp}}$ are vectors in the plane of the
sky and $\langle\rangle$ is an average over ${\bf r'_{\perp}}$.  The spatial
statistics of a magnetic field which can be approximated as a Gaussian random
variable are specified completely by the power spectrum of the field or,
equivalently, by its Fourier transform (the autocorrelation function).  If the
thermal electron density is uniform, then the RM structure function can be used
to infer the power spectrum of the magnetic field directly
\citep{EV03,laing08}. Even if these assumptions do not hold, the RM structure
function provides a robust way of characterizing the fluctuations of field and
density on different angular scales.

We computed RM structure functions at 4.0 and 1.3-arcsec resolution over the
areas shown in the top panels of Fig.~\ref{fig:sf}. These were chosen so that
the characteristics of the RM fluctuations are fairly homogeneous over each
sub-region, with enough independent points to determine reliable structure
functions. The sub-regions a, b, c and d (4.0\,arcsec) overlap with e, f, g and
h (1.3 arcsec), respectively. More points are blanked at the higher resolution,
so the precise selection of points is not the same. We have therefore not
attempted joint fits to the structure functions at the two resolutions, but the
parameters determined for corresponding sub-regions are in good agreement.
Sub-regions a and e, covering the N part of the E lobe, were defined to border
on but exclude the RM stripe; sub-regions b and f are positioned similarly in
the S of the lobe. We applied a
first-order correction for uncorrelated random noise by subtracting
$2\sigma^2_{\rm fit}$ from the structure function and averaged in logarithmic
separation bins. The results are plotted in Fig.~\ref{fig:sf}.

We fit the observed structure functions using a Hankel-transform method
\citep{laing08}. This exploits the fact that the observed RM image is closely
approximated by the convolution of the true RM image with the observing beam
provided that the wavelength is sufficiently small. In this case, the observed
structure function (including the effects of the beam) can be derived
straightforwardly for any assumed RM power spectrum. 
A cut-off power law (CPL) RM power spectrum: 
\begin{eqnarray}
\widehat{C}(f_{\perp}) &= & 0 ~~~~~~~~~~~~~~~~
f_{\perp}<f_{\rm min}\nonumber \\
                       &= & C_{0}f_{\perp}^{~-q} ~~~~~
f_{\rm min} \leq f_{\perp}\leq f_{\rm max}  \nonumber \\
                       &= & 0 ~~~~~~~~~~~~~~~~
f_{\perp}>f_{\rm max}\,.
\label{0755cpl}
\end{eqnarray}
where $f_\perp$ is a scalar spatial frequency, gives model structure functions
in good agreement with the data (Fig.~\ref{fig:sf}). 
We then estimated the maximum frequency (minimum
scale) by fixing these values and varying $f_{\rm max}$ to find the best match
to the mean value of $k$ for the region, also as described by \citet{laing08}.
Finally, we produced multiple realisations of RM distributions corresponding to the CPL model power spectrum on the observed grid, including the effects of the
convolving beam, and calculated their structure functions.
The rms of these
structure functions gives an estimate of the error due to statistical
fluctuations and is plotted as error bars attached to the observed points in
Fig.~\ref{fig:sf}, again as in \citet{laing08}.

The best-fitting parameters are quoted for each sub-region separately in
Table~\ref{07sffit}, together with the observed and predicted values of Burn
law $k$.

The RM fluctuations on large scales in the W lobe of 0755+37 are very weak. As a
consequence, the slopes of the model power spectra are not well constrained,
particularly at 4.0-arcsec resolution.
The best-fitting values of the minimum frequency $f_{\rm min}$ are
0.05\,arcsec$^{-1}$ close to the nucleus (regions c and g) and
0.03\,arcsec$^{-1}$ at larger distances (d and h). These correspond to
outer scales $\Lambda_{\rm max} = 1/f_{\rm min}$ of 16 and 30\,kpc,
respectively.  The best-determined slopes are for the E lobe at 1.3-arcsec
resolution, where $q = 2.3 \pm 0.2$ for region e (N of the source axis) and $q =
3.2^{+0.3}_{-0.4}$ for region f; the values for the overlapping regions a and b
at 4.0-arcsec resolution are consistent with these, but with larger errors.  The
fitted outer scale for region a is $\approx$150\,kpc, but most of the
large-scale power is in the linear gradient. After removing this, the slope is
unchanged, but the outer
scale for the residual fluctuations drops to $\approx$40\,kpc (Fig.~\ref{fig:sf}a).
Finally, there
is no firm evidence for a long-wavelength cut-off in the power spectrum of regions a
and
e (the Southern part of the E lobe), but the sampling on scales $\ga$50\,arcsec
($\ga$40\,kpc) is poor.

The maximum frequency can be chosen to make the predicted and observed values of
Burn law $k$ consistent in all sub-regions except b, e and f (E lobe), where the
former are too small for any choice of $f_{\rm max}$. In other words, an
extrapolation of the RM power spectrum to higher spatial frequencies predicts
too little power on scales $\la$1\,arcsec. This argues for an additional
depolarizing component, as we discuss further in Section~\ref{Depol}.

In Section~\ref{RMfluct}, we predict the variation of \srm across the source
using a single-scale model. The appropriate scale to use is roughly the magnetic
autocorrelation length $\lambda_B$ \citep{Murgia,laing08}, which we therefore
also give in Table~\ref{07sffit}. If no value of the minimum frequency could be
determined from the structure function, we assumed $f_{\rm min} =
0.005$\,arcsec, corresponding to separations slightly larger than the maximum in
the data for the regions in question. 

\begin{figure*}
\centering
\includegraphics[height=18cm]{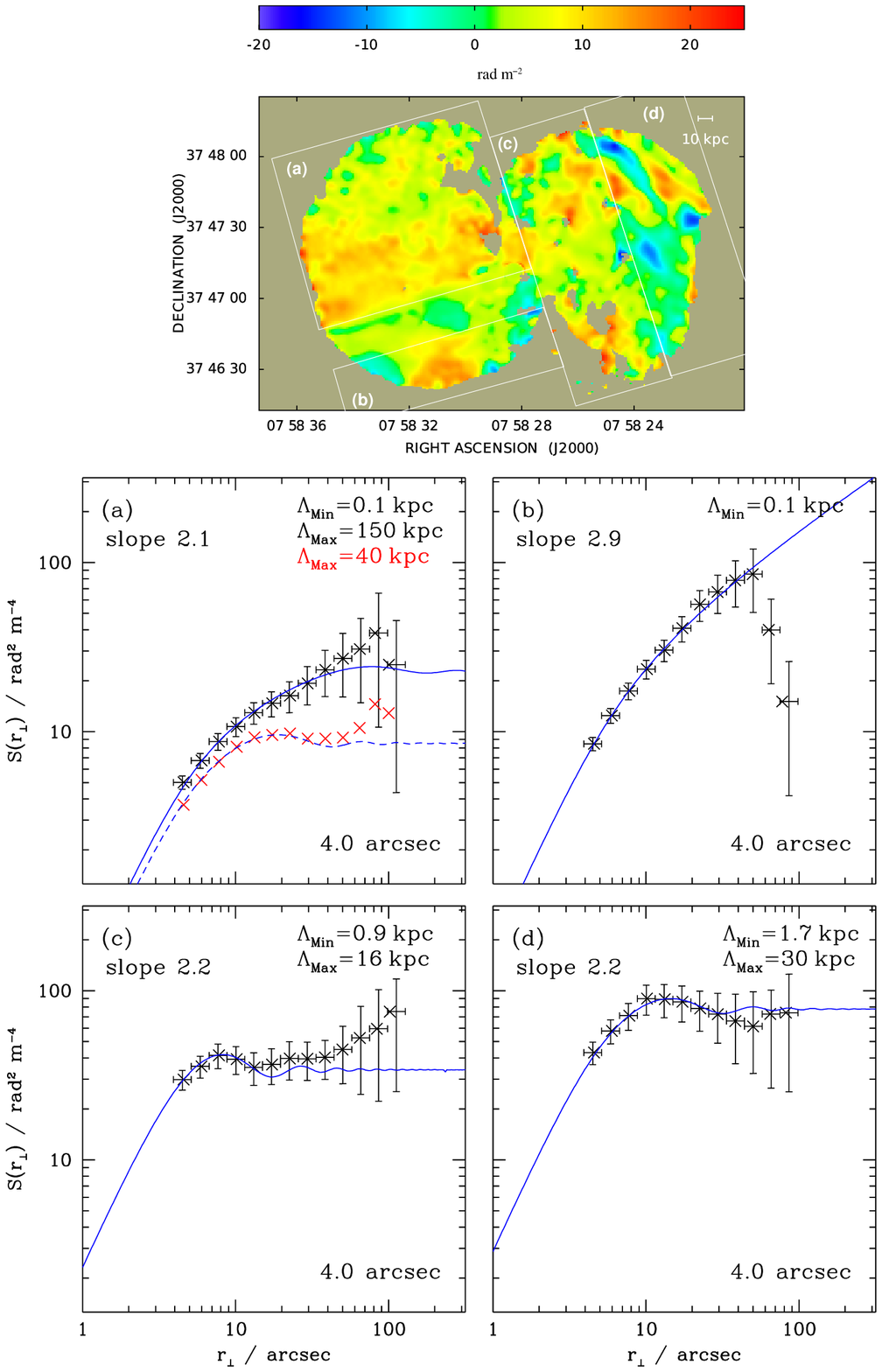}
\caption{Structure functions for the sub-regions of the RM map of 0755+37 
at 4.0\,arcsec FWHM as labelled in the RM image at the top.
The horizontal bars represent the bin widths and the crosses 
the mean separation for data included in the bins. 
The blue curves are the best fits, including the effects of the convolving beam. %
The error bars represent the rms variations for structure
functions derived from multiple realizations of the indicated power spectrum on the observed grid.
(a) Northern E lobe. (b) Southern E lobe. (c) Inner W lobe. (c) Outer W lobe.
In panel (a), the red crosses represent the observed structure function once
the linear RM gradient discussed in Section.~\ref{ongoingRM} is removed and the 
blue dashed line is the corresponding best fit.
\label{fig:sf}}
\end{figure*}

\begin{figure*}
\addtocounter{figure}{-1}
\centering
\includegraphics[height=18cm]{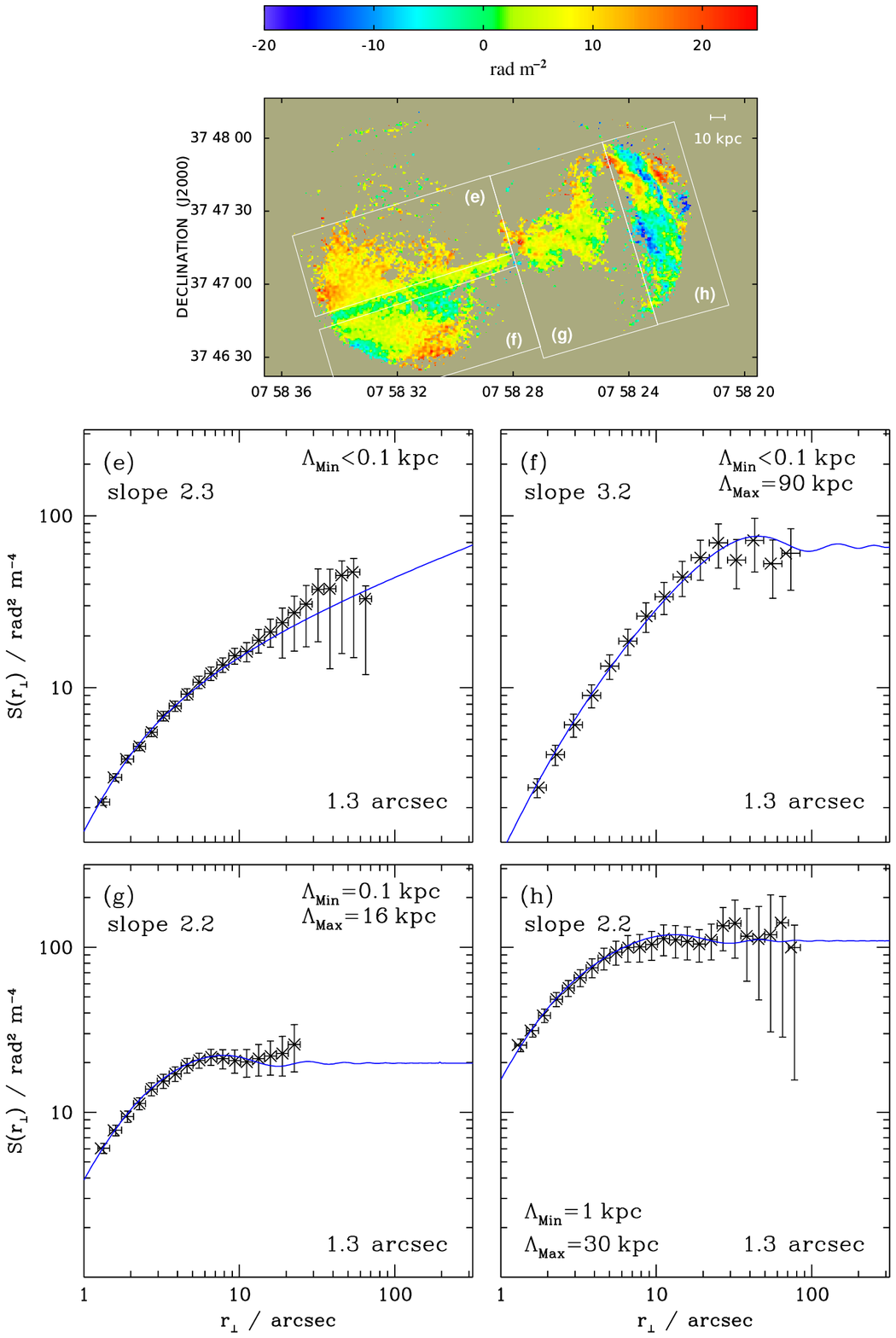}
\caption{(continued). Structure functions for the sub-regions of the RM map of
0755+37. The resolution is 1.3\,arcsec FWHM and the sub-regions are shown on the
map at the top.  (e) Northern E lobe. (f) Southern E lobe. (g) Inner W lobe. (h)
Outer W lobe.}
\end{figure*}

\begin{table*} 
\hspace{-30pt}
\caption{Best-fitting power spectrum parameters for the individual sub-regions
  of 0755+37. Col\,1: region, as shown in Fig.~\ref{fig:sf}. Col.~2: Beamwidth
  (FWHM, in arcsec). Col.~3: fitted power-law slope, $q$. If no lower error
  range 
  is quoted, $q$ is consistent with 0. Cols\,4 and 5: maximum and minimum
  spatial frequencies, in arcsec$^{-1}$. Cols\,6 and 7: observed and predicted 
  values of Burn law $k$, in rad$^2$\,m$^{-4}$. Col.~8: magnetic field autocorrelation 
  length, $\lambda_B$, in arcsec. $f_{\rm min} = 0.005$\,arcsec$^{-1}$ was assumed
  in the calculation of $\lambda_B$ if no value is quoted in Col.~5.   
  The second line for region (a) lists the best-fitting power spectrum parameters once
  the linear RM gradient discussed in Section.~\ref{ongoingRM} is removed.
}
\label{07sffit}
\centering      
\begin{tabular}{l l l r l r r r}  
\hline     
Region     & FWHM & \multicolumn{1}{c}{$q$} &\multicolumn{1}{c}{$f_{\rm max}$} & \multicolumn{1}{c}{$f_{\rm min}$} & \multicolumn{1}{c}{$k_{\rm obs}$} & \multicolumn{1}{c}{$k_{\rm syn}$} & \multicolumn{1}{c}{$\lambda_B$}\\
    & [arcsec] &  & [arcsec$^{-1}$] & [arcsec$^{-1}$] & [rad$^2$\,m$^{-4}$] & [rad$^2$\,m$^{-4}$] & [arcsec]   \\
       
\hline      
 a & 4.0 & 2.1$_{-0.4}^{+0.8}$ & 8.3 & 0.005 & 83$\pm$5 & 74$\pm$3 & 0.3 \\
   &     & 2.1$_{-0.4}^{+0.7}$ & 8.3 & 0.02  &          &          & 0.2  \\
 b & & 2.9$_{-0.3}^{+0.4}$ & $\geq$8.3  & & 99$\pm$3 & 25$\pm$3 & $\sim$3.5 \\
 c & & 2.2$^{+0.8}$ & 8.3   & 0.05 & 173$\pm$4 & 164$\pm$5 & 1.0\\
 d & & 2.2$^{+0.8}$ & 0.5  & 0.03 & 120$\pm$3 & 104$\pm$5 & 1.7\\
 e & 1.3 & 2.3$_{-0.2}^{+0.2}$& $\geq$8.3 &  & 56$\pm$4  & 14$\pm$5 & $\sim$0.5 \\
 f & & 3.2$_{-0.4}^{+0.3}$ & $\geq$8.3  & 0.01  & 70$\pm$4 & 11$\pm$5  & $\sim$6.0 \\
 g & & 2.2$_{-0.8}^{+0.5}$ & 8.3  & 0.05 & 70$\pm$5  & 55$\pm$5  & 0.8\\
 h & & 2.2$_{-0.9}^{+0.5}$ & 0.8  & 0.03 & 63$\pm$6 & 40$\pm$4 & 1.2  \\

\hline
\end{tabular}
\end{table*}

\section{Discussion}
\label{discuss}

Given the complexity of the RM distribution in 0755+37 and in particular its
obvious lack of axisymmetry, we have not attempted to build a three-dimensional
model (cf.\ \citealt{laing08,Guidetti10}).  Instead, we give a qualitative
discussion. We first (Section~\ref{RMfluct}) discuss the location and properties
of the magnetized plasma responsible for the resolved fluctuations in RM,
excluding the region to the N of the jet in the E lobe, to which we return in
Section~\ref{stripe}. The regions of enhanced depolarization around the jets are
the subject of Section~\ref{Depol}.

\subsection{RM fluctuations}
\label{RMfluct}

For 0755+37, the average value of \srm (5.9\,rad\,m$^{-2}$ at 1.3-arcsec
resolution) is comparable with that for NGC\,315 (also a member of a sparse
group; \citealt{Laing06}) and significantly less than for 3C\,31 and 3C\,449,
which are central galaxies in rich groups \citep{laing08,Guidetti10}. In turn,
much larger values are found for central radio galaxies in cool core clusters
such as M\,87, Hydra\,A and Cygnus\,A \citep{OEK,TP93,cyga}. The observed
correlation between \srm and X-ray surface-brightness \citep{Dolag01,Dolag06}
implies a scaling between central field strength and density $B_0 \propto
n_0^\eta$ with $\eta \approx 0.9$; 0755+37 is consistent with this relation. The
existence of a correlation between \srm and ambient density implies that the
plasma surrounding the radio source generates the Faraday rotation, but does not
differentiate models in which the dominant contribution is from undisturbed
IGM at large distances from those in which compressed material close to the
lobes is more important.

\begin{figure}
\centering
\includegraphics[width=8.5cm]{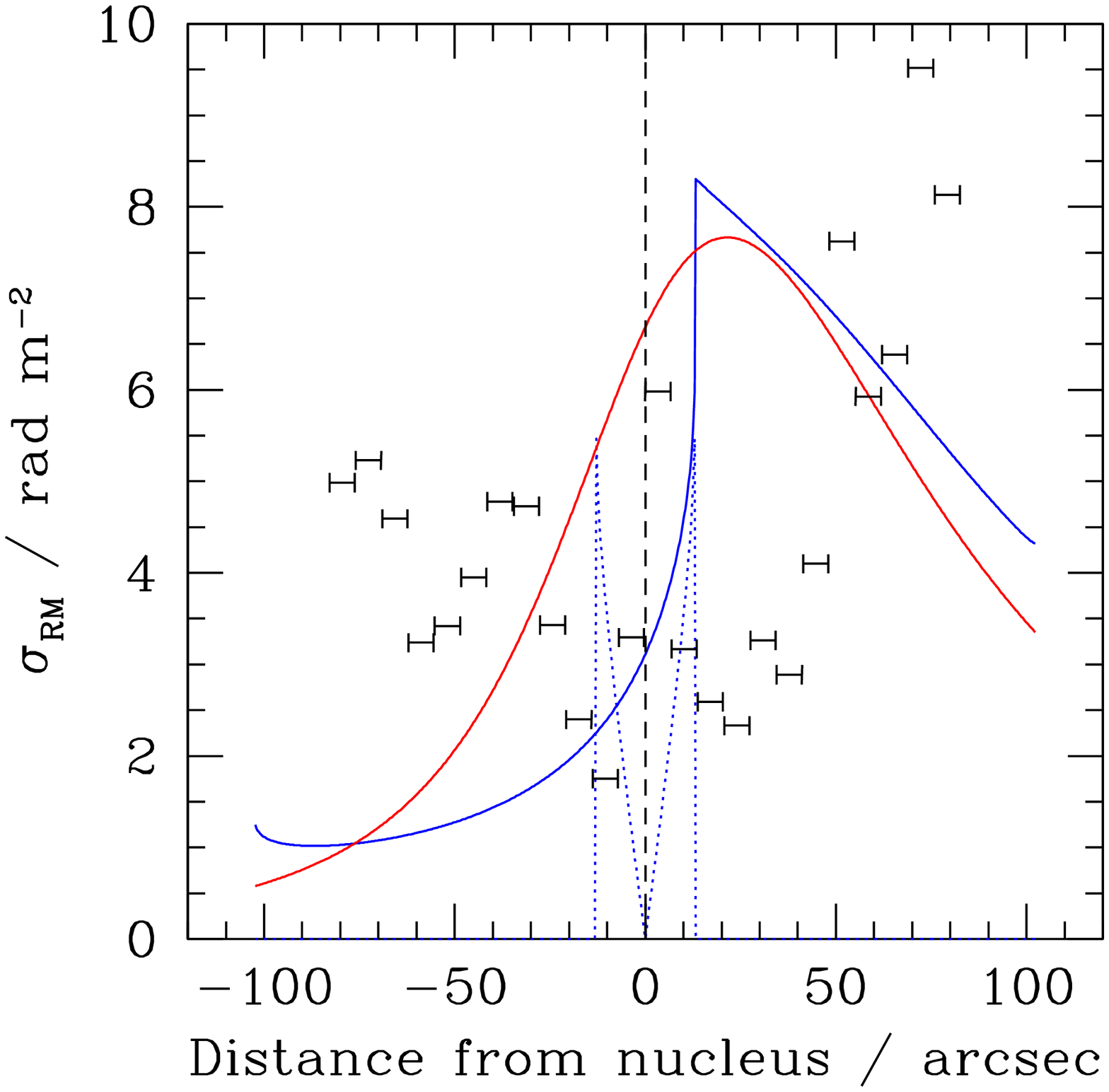}
\caption{Theoretical profiles of \srm along the source axis for 0755+37, calculated
  by numerical integration of a single-scale model, as in \citet{laing08},
  and superimposed on the observed \srm profile at 1.3-arcsec resolution from Fig.~\ref{0755kprof}(d). The assumed density
  distribution is the group beta model from Table~\ref{ongoingX}, scaled to
  electron density, and the
  magnetic field strength is $B \propto n^{0.5}$. The assumed scale size of the field fluctuations is 1\,kpc. Red
  curve: the source is taken to be a plane at an angle of 35$^\circ$ to the line
  of sight, with the approaching side on the left. The central field strength is
  $B_0 = 1.5$\,$\mu$G. Blue curves: thermal plasma
  is excluded from ellipsoidal cavities with semi-major and semi-minor axes of
  65 and 35\,kpc, respectively and $B_0 = 2$\,$\mu$G. The major axes of the cavities again make an
  angle of 35$^\circ$ with the line of sight. The cavity dimensions are chosen
  so that their projections on the plane of the sky approximately match the
  observed extent of the radio emission. The full line shows \srm along the line
  of sight to the front surface of the nearer lobe while the dotted line
  represents the contribution to \srm from material along the line of sight
  between the two lobes (this only affects emission from the receding lobe).
\label{fig:1scale}}
\end{figure}

A model in which the plasma distribution is spherically symmetrical, with the
field strength proportional to a power of the electron density, $B(r) \propto
n_e(r)^\eta$, $n_e(r)$ determined from X-ray observations and a planar source
model gives a good first approximation to the RM distribution expected from the
undisturbed IGM. Such a model was applied successfully to fit the RM
distribution of 3C\,449 by \citet{Guidetti10}. Slightly more elaborate
axisymmetric models in which thermal plasma is assumed to be excluded from
cavities occupied by relativistic, radio-emitting electrons, can also fit the
observed RM for 3C\,31 and Hydra\,A \citep{laing08,Kuchar11}. For 0755+37,
neither type of model can fit the observed \srm profiles adequately. In order to
demonstrate this, we show RM profiles along the source axis derived using the
single-scale approximation \citep{Felten96,laing08}.  The scale size is assumed
to be 1\,kpc (representative of the values in Table~\ref{07sffit}) and the
electron density profile is for the group component in Table~\ref{ongoingX},
scaled from the proton number density by the factor $n_e/n_p = 1.18$ appropriate
for a normal cosmic composition.  The central field strengths ($B_0 = 1.5$ and
$2$\,$\mu$G respectively for the planar and cavity models) are set so that the
maximum of the curve roughly matches the peak \srm at the end of the W lobe. For
a source modelled as a plane at an angle of 35$^\circ$ to the line of sight, the
RM profile is peaked at 45\,arcsec from the nucleus in the W lobe
(Fig.~\ref{fig:1scale}; red curve) and decreases smoothly away from this
point. In the more realistic model shown by the blue curve in
Fig.~\ref{fig:1scale}, thermal plasma is excluded from the radio lobes
(represented as ellipsoids). The \srm profile in the receding (W) lobe is quite
flat, while that for the approaching (E) lobe drops smoothly except for a slight
rise close to the end of the source.  The one-dimensional profiles 
are not strictly comparable with the observed results,
which are averaged across the lobes and depend on sampling details, but are
realistic enough to demonstrate a qualitative inconsistency: models in which the
density and field strength decrease monotonically away from the nucleus cannot
reproduce the peaks in \srm at the ends of the lobes, and in particular the
rapid rise at the extreme W of the source seen at 1.3-arcsec resolution
(Fig.~\ref{fig:1scale}). The implication is that the observed RM is not
dominated by the undisturbed group plasma surrounding the source, although this
may contribute at the level of \srm $\approx$\,2 -- 3\,rad\,m$^{-2}$ ($B_0
\approx 0.5 \mu$G). We have not included any prediction for the Faraday rotation
due to the central corona of hot plasma associated with the host galaxy
(Table~\ref{ongoingX}), since this will be localized within a few arcseconds of
the nucleus, but it may contribute to the observed peak in \srm at 1.3-arcsec
resolution (Fig.~\ref{fig:1scale}).

\begin{figure}
\centering
\includegraphics[width=8.5cm]{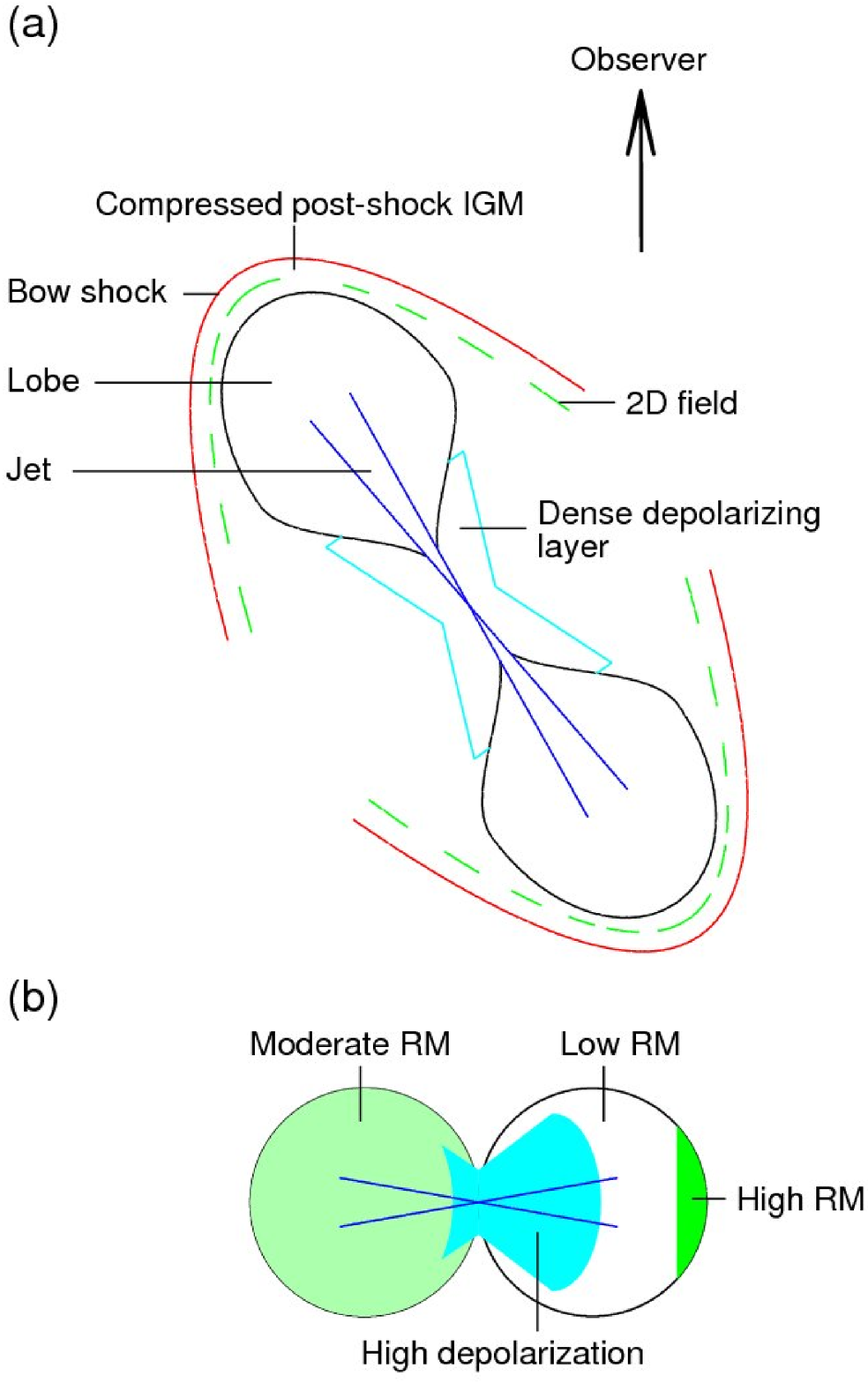}
\caption{Sketch of the proposed geometry for the location of magnetized thermal
  plasma around 0755+37. (a) Cross-section in a plane containing
  the source axis and the line of sight. The radio lobes (black), bow shocks
  (red), jets (blue) and post-shock magnetic field lines (green) are
  indicated, together with the location of the dense thermal plasma responsible
  for enhanced depolarization and the compressed IGM around the leading edges of
  the lobes. (b) Projection on the plane of the sky, showing the regions of high
  RM and depolarization corresponding to the magnetoionic plasma components
  sketched in panel (a). 
  \label{fig:sketch}}
\end{figure}

The most likely reason for the peaks in \srm at the ends of the lobes is that
the plasma immediately surrounding the fronts of the lobes has been compressed
by a bow-shock driven by the expanding radio source and that the post-shock
field has been both amplified and aligned in shells roughly parallel to the lobe
surface.  The geometry is sketched in Fig.~\ref{fig:sketch}. In this picture,
the maximum fluctuation amplitude occurs close to the end of the receding lobe,
where the magnetic field in the compressed IGM is preferentially aligned
parallel to the line of sight and the path length is high. If the shock is weak
or absent closer to the nucleus, then the remainder of the receding lobe will
have low RM values, since the line of sight does not pass through any compressed
IGM. In the approaching lobe, on the other hand, the compressed plasma covers a
larger fraction of the lobe, but the path length is relatively low and the field
is closer to the plane of the sky. We would therefore expect a moderate
level of RM fluctuation over a larger fraction of the lobe. This picture is in
at least qualitative agreement with the profiles of \srm observed in 0755+37
(Figs~\ref{0755kprof}c and d). Additional support for the idea comes from
simulations of FR\,II sources evolving in an isotropically-magnetized ICM
\citep[e.g.\ their Fig.~6]{huarte}, which show qualitatively similar RM distributions
to those we observe for model sources at 45$^\circ$ to the line of sight.
If we assume constant density and
two-dimensional field tangling in a plane containing the line of sight, then the
single-scale approximation gives $\sigma^2_{\rm RM} = (K^2/2) B^2 n_e^2 d
L$, where $d$ is the scale size and $L$ is the path length. Normalizing to
plausible values of $n_e$, $d$ and $L$, we find
\begin{eqnarray}
\frac{B}{\mu{\rm G}} &\approx& 0.35 \left (\frac{\sigma_{\rm RM}}{{\rm rad\,m}^{-2}}\right
  ) \left (\frac{n_e}{10^{-3} {\rm cm}^{-3}}\right )^{-1} \nonumber \\
&\times&\left (\frac{d}{1 {\rm
  kpc}}\right )^{-1/2}\left (\frac{L}{25 {\rm kpc}}\right )^{-1/2} \nonumber 
 \end{eqnarray}
At the leading edge of the W lobe, the RM fluctuations have $\sigma_{\rm RM}
\approx 8$\,rad\,m$^{-2}$ and the appropriate scale size is $\lambda_B \approx
1.2$\,kpc (Table~\ref{07sffit}, region h). For our fiducial path length and
density, the field strength would be $B \approx 2.5\mu$G.  Unfortunately the {\sl XMM-Newton}
count levels in the outer parts of the 0755$+$37 group are too low to identify
any shells of high density surrounding the leading edges of the lobes.  A more sensitive X-ray
observation would be required to search for enhancements associated with
regions of high \srm. 

This simple picture is complicated by the evidence for a preferred direction in
the RM distribution at the end of the W lobe (Fig.~\ref{fig:wlobe}). As noted
earlier, the anisotropic RM features resemble the `bands' found in four other
lobed sources \citep{Guidetti11}.  The straightness of the best-defined bands and their
orthogonality to the source axis led \citet{Guidetti11} to argue that they must be produced by
two-dimensional magnetic fields draped around the leading edges of the lobes in
the post-shock IGM, rather than by simple compression.  It
is more straightforward to produce shorter, misaligned arc-like RM features like those seen in
0755+37 by compression of an ordered pre-shock field \citep[Fig.10]{Guidetti11},
but generating reversals in RM is more problematic.

The conditions under which the RM and magnetic-field power spectra have the same
form (i.e.\ that the density and path-length are uniform over an area large
compared with the fluctuation scale and that the field can be taken to be an
isotropic Gaussian random variable) are clearly not satisfied for 0755+37.  We
note, however, that all of the derived indices are relatively flat, with
with $q$ in the range 2.1 to 3.2, as we have found for other sources
\citep{laing08,Guidetti10,Guidetti11} and
that none of the well-constrained slopes are consistent with the Kolmogorov 
value of $q = 11/3$ expected if the turbulence is hydrodynamic \citep{Kol41} or for an MHD cascade \citep{GS97}.  One interesting possibility, consistent with the field geometry sketched in Fig.~\ref{fig:sketch}, is that the magnetic
turbulence is closer to two-dimensional, in which case $q = 8/3$ \citep{MS96}. The range of 
outer scales we find from the RM structure functions for 0755+37 bracket the value of 65\,kpc determined for the magnetic field power spectrum in 3C\,449 by \citet{Guidetti10}.
Their analysis includes the effects of large-scale spatial 
variations in density and field strength and is therefore more rigorous
than we can attempt here.   
Further investigation
of the spatial statistics of the field around 0755+37 would again require a much
better understanding of the density distribution and geometry of the source.

\subsection{Anomalous RM in the E lobe}
\label{stripe}

The simple picture presented in Section~\ref{RMfluct} ignores the region with a
linear RM gradient and low \srm which forms the Northern part of the E
lobe. There is a hint that this is associated with a deficit of surrounding
plasma: the {\sl XMM-Newton} image (Fig.~\ref{0755X}) shows a lower level of
X-ray emission coincident with this part of the lobe. As noted in
Section~\ref{07x}, however, this feature is only significant at the 2$\sigma$
level. There is also some evidence from radio imaging for continuing, but less
well collimated flow after the termination of the E jet. Images of the spectral
index between 1385 and 4860\,MHz \citep[Figs\,6e and f]{laing11} show
flat-spectrum ($\alpha \approx 0.7$; $S(\nu) \propto \nu^{-\alpha}$) emission
extending due North from the end of the E jet to the Northern boundary of the
lobe. It therefore seems possible that the flow deflects Northwards into a
region of lower external density.

The `RM stripe' (Fig.~\ref{rmstripe}) appears to mark an abrupt boundary
between fluctuating and ordered RM and lies close to a continuation of the
jet axis 
(Fig.~\ref{0755X}). Its RM is +2\,rad\,m$^{-2}$, significantly lower than our
best estimate of the Galactic value ($\approx$ +8\,rad\,m$^{-2}$). North of the
stripe, the RM increases over a short distance to +10\,rad\,m$^{-2}$, the
fluctuation amplitude suddenly drops and the smooth gradient in RM begins
(Fig.~\ref{tran_prof}).  The nature of the stripe remains a mystery.

The origin of the RM gradient in 0755+37 is also obscure. The smoothness of the
variation, together with the lack of small-scale fluctuations, argues that the
field responsible for the Faraday rotation is well-ordered on scales of several
tens of kpc and that the path length also varies slowly.  A clear linear trend
in RM is also seen over $\approx$700\,arcsec (235\,kpc at the source) along the
jets of NGC\,315, which is another source located in a sparse environment
\citep{Laing06}.  This compares with 70\,arcsec (58\,kpc) for 0755+37. The lack
of dependence on local gas density and the relatively large value of the
Galactic RM led \citet{Laing06} to argue that the gradient in NGC\,315 is due to
our Galaxy. Both of these arguments are much weaker for 0755+37, where the
gradient (which in any case covers only part of the source) is over a much
smaller angular scale and the absolute value of the Galactic contribution is a
factor of $\approx$10 smaller.  The component of the gradient parallel to the
jets in NGC\,315 is 0.075\,rad\,m$^{-2}$\,kpc$^{-1}$ at the source and the (much
less well-determined) total value is 0.16\,rad\,m$^{-2}$\,kpc$^{-1}$. For
0755+37, the gradient, 0.14\,rad\,m$^{-2}$\,kpc$^{-1}$, is comparable in
magnitude. It remains unclear whether the gradients in these two sources have
similar origins.  Comparable linear RM gradients in other sources might well be
masked by larger fluctuation levels.

\subsection{Depolarization}
\label{Depol}

The high values of Burn law $k$ observed around the edges of both jets
(particularly the W one) in 0755+37 are not associated with particularly large
RM fluctuation amplitudes (Figs~\ref{0755RMkfig}a and c).
They appear to result
from thermal plasma which introduces depolarization without much ordered Faraday
rotation. We conjecture that there is a layer of dense thermal plasma, forming a
rough hollow cone,  
immediately surrounding the jets and the inner lobes and containing a
magnetic field with a small mean component and many reversals across the
observing beam. This is the type of model described by \citet[Eq.~21]{Burn66},
which again has $p(\lambda) \propto \exp(-k\lambda^4)$, but this time with zero
RM in the limit of many reversals. The proposed geometry (sketched in
Fig.~\ref{fig:sketch}) has the following implications.
\begin{enumerate}
\item Emission from the inner parts of both lobes is affected by the dense
  depolarizing layer, but there is more depolarization in the half of the
  receding (W) lobe closer to the nucleus, primarily because a larger fraction
  of the lobe emission is seen through the dense, thermal plasma.
\item There is a minimum in the depolarization on-axis (at the locations of the
  jets) since the path length is lower there than around the edges.
\item In the suggested model, $k = 2K^2n_e^2B^2dL$, where $L$ is the path length
  through the depolarizing material and $d$ is the scale size of field
  reversals \citep{Burn66}. 
  $K = 811.9$ for $n_e$ in cm$^{-3}$, $B$ in $\mu$G
  and distances in kpc 
  (equation~\ref{equarm}). 
\item If this material is in a hollow cone, then $L \propto R$, where $R$ is the
  distance from the core. The slow variation of $k$ with $R$ (Fig.\ref{0755RMkfig}c)
  suggests that the product $nB$ must decrease roughly $\propto R^{-1/2}$.
\item A typical value of $k = 240$\,rad$^2$\,m$^{-4}$. This gives 
$B/\mu{\rm G} \approx 20 (d/0.1{\rm kpc})^{-1/2}(L/5{\rm
  kpc})^{-1/2}(n_e/10^{-3}{\rm cm}^{-3})^{-1}$, where we have parametrized the
  unknown quantities by ratios to plausible values. $n_e$ and $L$ could be
  estimated from high-resolution X-ray observations. 
  \end{enumerate}
Similar depolarized areas observed in M\,84 and 3C\,270 \citep{Guidetti11} are
associated with dense, highly disturbed X-ray emitting plasma around the jets
and inner lobes \citep{M84Chandra,Worr10}.  We have searched for similar
structures in our {\sl XMM-Newton} image (Fig.~\ref{0755X}) without success: the
resolution is too low to distinguish them from the bright core, jet and corona in
the centre of the source or from more diffuse emission on larger scales.

The excess of observed over predicted Burn law $k$ in sub-regions b, e and f of the E
lobe (Table~\ref{07sffit}) may be due to the inclusion of emission seen through
the depolarizing layer. In general, the presence of the layer makes it difficult
to constrain the minimum scale of the medium responsible for larger-scale
fluctuations in Faraday rotation where the two overlap and the values of $f_{\rm
  max}$ given in Table~\ref{07sffit} should be treated with caution.  

\subsection{Implications for the interpretation of depolarization asymmetry}
\label{LG}

It has been known for many years that the lobe containing the brighter (or only)
jet is less depolarized than that on the counter-jet side for both FR\,I and
FR\,II sources \citep*{L88,Garri88,GCL91,Morg97}. This is interpreted
straightforwardly as an orientation effect \citep{L88}: if the jets are
symmetrical and relativistic, then the approaching one will appear brighter as a
result of Doppler beaming, and will also be seen through less Faraday-rotating
material, resulting in lower depolarization or RM dispersion, depending on the
resolution. The simplest models of the effect assume that the magnetized medium has
spherical symmetry and the lobes have negligible depth \citep{GC91}, but
these approximations are not always adequate.
The present study has prompted us to clarify the nature of the difference
in Faraday depth between the approaching and receding sides for more realistic
distributions of field and thermal plasma.  For 0755+37, we
have identified four different variants, as follows.
\begin{enumerate}
\item Faraday rotation produced by a varying field in the undisturbed IGM
  surrounding the host galaxy will have a centrally-peaked \srm profile. This is
  asymmetric unless the source is in the plane of the sky. We expect thermal
  plasma to be excluded from the radio lobes: if these occupy a significant
  volume, then the profile will be modified even if the external density and
  field are unaltered (e.g.\ as in Fig.~\ref{fig:1scale}; more realistic
  simulations are described by \citealt{laing08}).
\item The IGM will be compressed by the expansion of the radio lobes,
  particularly if this is supersonic, in which case a bow shock forms. The
  strongest compression will be around the leading edges of the
  lobes. Components of an external field perpendicular to the shock will be
  amplified, leading to an anisotropic post-shock configuration. Faraday
  rotation produced by the shocked IGM will be higher for the receding lobe not
  only because the path is longer, but also because the field is on
  average closer to the line of sight. For some geometries, the largest RM
  fluctuations in the receding lobe will be restricted to a small area close to
  the end of the source (Fig.~\ref{fig:sketch}).  
\item The simple compression model cannot entirely explain the anisotropic RM
  structures seen in the receding lobe of 0755+37 and fails completely for
  sources showing RM bands across the full widths of their lobes. 
  As pointed out by \citet{Guidetti11}, the bands require a two-dimensional 
  `draped' field configuration. It is not yet clear whether the amplitude 
  of the bands is systematically lower for the approaching lobe, although this 
  is the case for the one known example with a reasonably small angle to the 
  line of sight, 0206+35 \citep{Guidetti11}.  
  If the effect is general, then the likely cause is
  a combination of differences in path length and field orientation through 
  the draped-field region. 
\item There is enhanced depolarization associated with shells of denser thermal
  plasma around the inner lobes and jets of M\,84 and 3C\,270 \citep{Guidetti11}
  and we have suggested that this is also the case for 0755+37. There is little
  ordered Faraday rotation associated with the depolarization, indicating that
  the field has many reversals across the beam and/or that the plasma is clumped
  on small scales.  Differential depolarization between the approaching and
  receding lobes is produced primarily because a larger fraction of emitting
  volume of the latter is seen through the depolarizing plasma
  (Fig.~\ref{fig:sketch}).
\end{enumerate}
All four of these geometries produce more Faraday rotation of radiation from the
receding lobe; which is dominant will depend on the type of radio source and the
state of the thermal plasma associated with the galaxy and group or cluster.

\section{Summary}
\label{concl}

We have derived detailed images of Faraday rotation measure (RM) and
depolarization (quantified by fitting to a Burn law) at resolutions of 4.0 and
1.3\,arcsec FWHM for the FR\,I radio galaxy 0755+37 from deep VLA observations
at 1385, 1465 and 4860\,MHz.  The spatial statistics of the RM distributions
have been investigated using a structure-function technique. We have also
presented new {\sl XMM-Newton} observations of X-ray emission from 0755+37.

The results of this analysis can be summarized as follows.
\begin{enumerate}

\item

The X-ray emission from hot gas on scales comparable with the size of the associated sparse
group of galaxies can be fit by a beta model with central proton density 
$1.2 \times 10^{-3}$\,cm$^{-3}$, core radius 66\,kpc, $\beta = 0.7$ and
temperature kT = 1.3\,keV.

\item

The polarization position angles accurately follow a $\lambda^2$ relation and
the observed depolarization is low. This is consistent with Faraday rotation by
foreground material, as found for many other sources, but the range of rotation
is too small to require it in this particular case.

\item

The amplitude of the RM fluctuations in 0755+37 (\srm = 5.9\,rad\,m$^{-2}$
averaged over the source) is smaller than for FR\,I sources embedded in rich
groups of galaxies (\citealt{laing08,Guidetti10,Guidetti11}) and comparable with
that in NGC\,315, which is also in a sparse environment \citep{Laing06}.
This is consistent with the known correlation between gas density and \srm
\citep{Dolag06}.

\item 

There is an asymmetry in RM structure between the E (approaching) and W
(receding) lobes: the N part of the E lobe is characterized by a positive RM
with a linear gradient roughly transverse to the source axis whereas the S part
of that lobe and the whole of the W lobe show irregular fluctuations. There is a
hint that the linear gradient is associated with a lower external density.

\item

The boundary between smoothly-varying and patchy RM in the E lobe is marked by a
narrow `stripe' of low and constant RM. This is almost aligned with the jet
axis. It covers the outer part of the jet, but also continues beyond the end of
the jet to the edge of the lobe. Its cause is not understood.

\item 

The largest RM fluctuations are at the end of the W (receding) lobe. They show
reversals with respect to the Galactic value and are anisotropic, with
iso-RM contours preferentially parallel to the edge of the lobe and to a
prominent brightness step. The anisotropic features resemble RM bands
\citep{Guidetti11}.

\item

The distribution of \srm cannot be explained by a model in which a
tangled magnetic field is embedded in a spherically-symmetric distribution of
external plasma, even if the effect of excluding the plasma from the radio lobes
is taken into account.  Instead, we suggest that the resolved Faraday-rotation
fluctuations are dominated by compressed IGM around the ends of the
lobes. Unfortunately, the
X-ray surface brightness is too low for us to test for density enhancements directly.
We
point out that the magnetic field in the compressed plasma will be orientated
preferentially along the line of sight in front of the receding lobe,
contributing to the observed asymmetry in Faraday rotation. The geometry and
density distribution cannot be determined well enough to allow us to estimate field
strengths accurately, but values of a few $\mu$G are plausible.

\item

The structure functions in different parts of the source can be fit
assuming that the RM power spectra are cut-off power laws. The fits show
that there is little power on spatial scales $\ga$16 -- 30\,kpc in the W
lobe and $\ga$90 -- 150\,kpc in the E lobe.  We find
that the slopes of the power laws are in the range from 2.1 to 3.2 where they
can be determined adequately, inconsistent with the value of 11/3 expected for
Kolmogorov turbulence, but similar to estimates for other sources
(e.g.\ \citealt{laing08,Guidetti10,Guidetti11}). 
The fitted minimum scales range from $\approx$1\,kpc to $\la$0.1\,kpc.

\item

Excess depolarization is observed at the edges of both jets, particularly in the
receding lobe. There is no obvious correlation with RM. We suggest that this is
due to a layer of denser plasma around the jets and inner lobes with a magnetic
field tangled on small scales. 

\end{enumerate}

The lack of symmetry in the source structure precludes three-dimensional
modelling which, together with accurate estimation of field strengths, must
await a more detailed understanding of the external density distribution. This
requires more sensitive and higher-resolution X-ray observations.

\section*{Acknowledgements}

We thank Martin Huarte-Espinosa and Martin Krause for useful clarifications of
their work, Matteo Murgia for the use of the Synage package and Martin
Hardcastle for use of his MCMC surface-brightness profile modelling software.
RAL is grateful to INAF-IRA for hospitality.  DG thanks Lorenzo for delaying his
arrival until the completion of this work. The National Radio Astronomy
Observatory is a facility of the National Science Foundation operated under
cooperative agreement by Associated Universities, Inc. JHC would like to
acknowledge support from the South East Physics Network (SEPNet).

\bsp

\label{lastpage}

\end{document}